\documentclass[final,leqno,onefignum,onetabnum]{siamltex1213}
\usepackage{amssymb, amsmath, latexsym, amssymb, amsfonts, amsbsy}
\usepackage{extarrows}

\def\pI{{\mathbf{I}}}
\def\pN{{\mathbf{N}}}
\def\pA{{\mathbf{S_A}}}
\def\pC{{\mathbf{S_C}}}
\def\tx{{\mathbf{x}}}
\def\tm{{\mathbf{m}}}
\def\ttn{{\mathbf{n}}}

\newcommand{\ud}{\mathrm{d}}
\newcommand{\tr}{\mathrm{tr}}
\newcommand{\reffig}[1]{Figure \ref{#1}}
\newcommand{\refequ}[1]{(\ref{#1})}

\title{On a molecular based Q-tensor model for liquid crystals with density variations}

\author{Song Mei\thanks{School of Mathematical Sciences, Peking University, Beijing, 100871, P.R. China. 
(\email{meisong541@pku.edu.cn}). Current affiliation: Institute for Computational and Mathematical Engineering, Stanford University, CA, 94305. (\email{songmei@stanford.edu})} \and Pingwen Zhang\thanks{LMAM, CAPT and School of Mathematical Sciences, Peking University, Beijing, 100871, P.R. China. (\email{pzhang@pku.edu.cn)}}}

\begin{document}
\maketitle 
\slugger{mms}{xxxx}{xx}{x}{x--x}

\begin{abstract}
In this article, we study the new Q-tensor model previously derived from Onsager's molecular theory by Han \textit{et al.} [\textit{Arch. Rational Mech. Anal.}, 215.3 (2014), pp. 741-809] for static liquid crystal modeling. Taking density and Q-tensor as order parameters, the new Q-tensor model not only characterizes important phases while capturing density variation effects, but also remains computationally tractable and efficient. We report the results of two numerical applications of the model, namely the isotropic--nematic--smectic-A--smectic-C phase transitions and the isotropic--nematic interface problem, in which density variations are indispensable. Meanwhile, we show the connections of the new Q-tensor model with classical models including generalized Landau-de Gennes models, generalized McMillan models, and the Chen-Lubensky model. The new Q-tensor model is the pivot and an appropriate trade-off between the classical models in three scales.
\end{abstract}

\begin{keywords}
Liquid crystals, Q-tensor model, density variations, smectic phase, phase transition, isotropic-nematic interface\end{keywords}

\pagestyle{myheadings}
\thispagestyle{plain}
\markboth{A Molecular Based Q-tensor Model for Liquid Crystals}{A Molecular Based Q-tensor Model for Liquid Crystals}

\section{Introduction}
A proper choice of order parameters is the most important perspective for building a sound model for liquid crystals. The order parameters should be simple and explanatory in terms of mathematics and physics, while efficient in computations. From the viewpoint of theoretical analysis, it is desirable to adopt order parameters that are as simple as possible in order to capture essential phenomena of liquid crystals, with four phases the most important: isotropic($\pI$), nematic($\pN$), smectic-A($\pA$), and smectic-C($\pC$) phases. As for computational aspects, the order parameters should be discretely representable with reasonable dimensions while keeping the energy functional well-posed, concise, and efficient.

The classical models for static liquid crystals can be classified into three scales: molecular models, tensor models, and vector models. Each of these models have merits and limitations in respect to mathematics, physics and computations.

\subsection{Review of Previous Models}
The molecular models for liquid crystals are based on microscopic statistical physics. By means of the cluster expansion, Onsager \cite{model: Onsager} pioneered the molecular field theory, in which the order parameter is the density function for the position and orientation of molecules. Onsager's molecular model is established on sound physical principles, and contains no adjustable parameters. However, the molecular model is not clearly related to macroscopic properties, and the high dimension of the order parameter imposes considerable obstacles in computations. Based on Onsager's molecular theory, the McMillan model \cite{model: McMillan} and a list of molecular models  \cite{model: molecular hard core} \cite{model: molecular potential} \cite{model: P O} parameterized the density function by some scalar order parameters to model the smectic phases. By lowering the dimensions, these model are very efficient in computations. However, the order parameters of these models are spatially invariant, so that they cannot model detailed physical phenomena with confined geometry and spatial variance. 

On the other hand, a vector model for liquid crystals was phenomenologically proposed by Oseen \cite{model: O-F}, where the order parameter is a vector field $\ttn(\tx)$ representing the director. As a development of Landau-Ginzburg theory and Oseen-Frank theory, Chen and Lubensky \cite{model: chen} introduced the density as another order parameter to characterize $\pN$-$\pA$-$\pC$ phase transitions. The Chen-Lubensky model is famous for its conciseness. The internal coefficients can be measured through experiments. However, this model presumes liquid crystals to be uniaxial, and the director $\ttn$ is singular at defect points, which renders it difficult to characterize some small scale phenomena, such as defects and interfaces.

Overcoming the drawbacks of the molecular models and the vector models, the well-known Landau-de Gennes model \cite{model: de Gennes} was proposed with the energy functional
\begin{equation}
\label{ldg}
\begin{aligned}
F^{LG}[Q]&=\int_\Omega\left( \frac{A(T-T^*)}{2}\tr(Q^2)-\frac{B}{3}\tr(Q^3)+\frac{C}{4}(\tr(Q^2))^2\right) \ud \tx\\
&+\int_\Omega\left( L_1 \vert \nabla Q \vert^2+L_2 \partial_j Q_{ik}\partial_k Q_{ij}+L_3 \partial_j Q_{ij} \partial_kQ_{ik}+L_4 Q_{lk} \partial_k Q_{ij} \partial_l Q_{ij}\right)  \ud \tx,
\end{aligned}
\end{equation}
where $A$, $B$, $C$, and $L_i$ are constants. The order parameter is a $3 \times 3$ symmetric traceless tensor field $Q(\tx)$, which is the second moment of molecules' orientation at every point. 

The Q-tensor is a desirable order parameter providing information on both the preferred molecular orientation and the degree of orientational order at every given point, while capturing essential physical properties. Various phenomena were studied with the Landau-de Gennes model, e.g. phase transitions in confined geometries \cite{LD: geometry1} \cite{LD: geometry2}, wetting phenomena \cite{LD: wet1} \cite{LD: wet2}, surface-induced bulk alignment \cite{LD: surface1} \cite{LD: surface2}, and defects and disclinations \cite{LD: defect1} \cite{LD: defect2}. 

The order parameter and the energy functional of the Landau-de Gennes model are also simple enough to perform rigorous mathematical analysis. Ball and Majumdar proved that the energy of Landau-de Gennes model is unbounded from below as $L_4 \neq 0$, and proposed to modify the entropy term from a polynomial into a thermotropic one in order to avoid the unboundedness \cite{paper: jball}. Furthermore, Ball and Zarnescu proved that for simply-connected domains and in Sobolev space $W^{1,2}$ with corresponding boundary conditions, the Landau-de Gennes theory and the Oseen-Frank theory coincide \cite{orient}.

Various generalizations of the Landau-de Gennes model \cite{model: P O} \cite{model: GLGT} \cite{model: Mukherjee} \cite{model: B C T} were proposed to include smectic phases. The work of Pajak and Osipov \cite{model: P O} is a generalization of the McMillan model \cite{model: McMillan} and the Landau-de Gennes model, which starts from the self-consistent field theory and adopts the one mode approximation to parameterize the density function. The details of this model are provided in section \ref{section: GMM}. This model is efficient in computations, but the order parameters are spatially invariant, failing to characterize some physical phenomena with confined geometry and spatial variance. The works of Mukherjee \cite{model: Mukherjee} and Biscari \textit{et al.} \cite{model: B C T} are also generalizations of the Landau-de Gennes theory. The Q-tensor is coupled with the complex smectic order parameter, and the spatial inhomogeneity of the order parameters enables the model to characterize smectic phases. However, it is noteworthy that these models did not explain how the energy functionals were derived in details.

\subsection{A Recent Model}
In a recent paper by Han \textit{et al.} {\cite{HLWZ}}, a systematic way of modeling static liquid crystals with uniaxial molecules was proposed. To be more precise, starting from Onsager's molecular theory, a new Q-tensor model was presented incorporating the Bingham closure and a Taylor expansion with truncation at low order moments. The coefficients in the new Q-tensor model were approximated in terms of the microscopic shape factor $\varepsilon = D/L$ by assuming the interaction potential to be the volume exclusion potential of rigid rod-like molecules. Here, $D$ is the diameter of the semisphere at the ends of the rods, and $L$ is the length of the rods. In modeling the nematic phase, the Oseen-Frank model and the Ericksen model were derived using the new Q-tensor model by assuming a constant density. Three elastic constants $K_1$, $K_2$, and $K_3$, measuring the strains on liquid crystals in deformation, were calculated analytically.  In modeling the smectic phase, under the uniaxial assumption, some preliminary numerical results regarding $\pI$-$\pN$-$\pA$ phase transitions were presented.

The energy functional of the new Q-tensor model in a modified version reads
\begin{equation}\label{energy0}
\begin{aligned}
F[c, Q&]=F_{bulk} + F_{elastic, 2}+ F_{elastic, 4},\\
\end{aligned}
\end{equation}
where
\begin{align}
\nonumber\beta F_{bulk}=&\int_\Omega c(  \tx) \Big(\ln c( \tx) + B_Q:Q-\ln Z( \tx) \Big) \ud \tx\\
\label{bulk}+&\frac{1}{2}\int_\Omega \Big[A_{1}c^2 - A_{2} \vert cQ \vert ^2 - A_{3}\vert cQ_4\vert ^2\Big] \ud \tx,\\
\nonumber\beta F_{elastic, 2}=& \frac{1}{2}\int_\Omega\Big[-G_1 \vert \nabla c\vert^2 +G_2 \vert \nabla(cQ)\vert^2+G_3\partial_i(cQ_{ik})\partial_j(cQ_{jk})  -G_4 
\partial_i(cQ_{ij})\partial_j(c)\\
\label{elastic2}&+ G_5\vert \nabla(cQ_4)\vert^2+G_6\partial_i(cQ_{4iklm})
\partial_j(cQ_{4jklm})+G_7 \partial_i(cQ_{4ijkl})\partial_j(cQ_{kl})\Big] \ud \tx,\\
\nonumber\beta F_{elastic, 4}=&\frac{1}{2}\int_{\Omega}\Big[H_1\vert \nabla^2 c\vert^2+ H_2 \vert \nabla^2 (cQ)\vert^2 \\
\label{elastic4}&+ H_3\partial_{ij}(cQ_{ij})\partial_{kl}(cQ_{kl}) + H_4 \partial_{ik}(cQ_{ip})\partial_{jk}(cQ_{jp}) \Big] \ud \tx.
\end{align}
Here, $\beta=\frac{1}{k_BT}$ is the thermodynamic beta. $c(\tx)$ is the density. Q-tensor $Q(\tx)$ is the traceless second moment of the orientational variable $\tm$, and $Q_4(\tx)$ is the traceless fourth moment of the orientational variable $\tm$. $B_Q$ is the Bingham distribution parameters, and $1/Z(\tx)$ is the normalizing constant for the Bingham distribution. The bulk energy $F_{bulk}$ contains the entropy and the quadratic terms of the order parameters. The second order elastic energy $F_{elastic, 2}$ contains the derivative terms of the order parameters; the fourth order elastic energy $F_{elastic, 4}$ contains the second order derivative terms of the order parameters. Note that here the fourth order elastic energy $F_{elastic, 4}$ is truncated: only the positive definite terms are preserved in order to ensure the lower boundedness of the energy functional. The detailed descriptions of the energy functional are presented in the main body of this paper.

\subsection{Our contributions}
In this paper, regarding the new Q-tensor model as a phenomenological model, we focus on its density variation effects, and show its effectiveness in terms of mathematics, physics, and computations. The introduction of the density $c(\tx)$ as another order parameter exhibits many benefits. It empowers the model to characterize the smectic phases, and to characterize the density variations in small scale phenomena such as defects and interfaces. Since a few Fourier modes are enough to characterize the profiles of the order parameters, the model can be solved with low computational costs. Numerical experiments are performed to study $\pI$-$\pN$-$\pA$-$\pC$ phase transitions, and the results are compared with experimental results. The $\pI$-$\pN$ interface problem is also studied in the new Q-tensor model setup, where the numerical results are compared with previous theoretical results and experimental results. 

\begin{figure}
\centering
\includegraphics[width=0.9\linewidth]{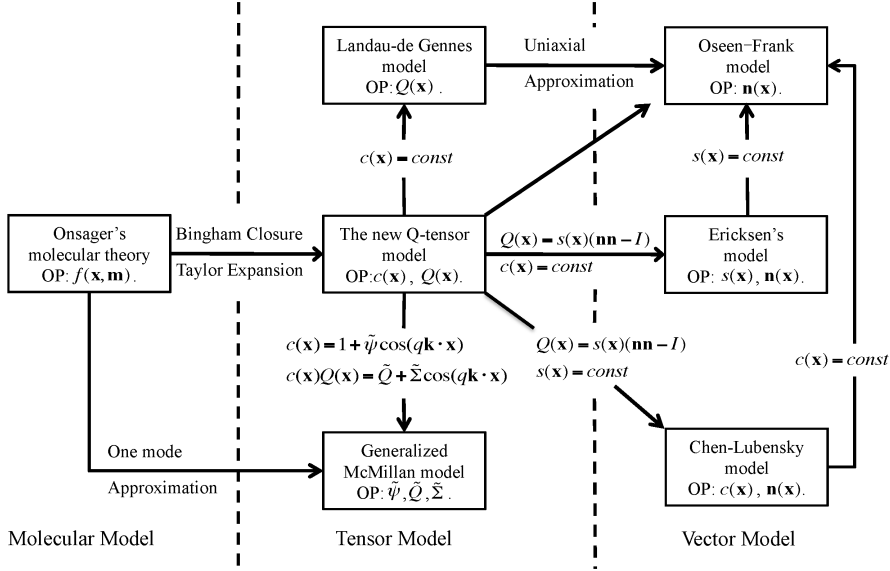}
\caption{Three-scaled schema in static liquid crystals modeling. OP is short for order parameters.}\label{relations}
\end{figure}

In addition, we will elaborate on the strong connections of the new Q-tensor model with the classical models of liquid crystals in three scales. Derived from Onsager's molecular theory, the new Q-tensor model can generate all the other classical models with some assumptions and approximations. \reffig{relations} shows the classical models in three scales and their relations. 

\subsection{Outline}
This paper is organized as follows. In section \ref{section: prop}, we demonstrate the mathematical properties of the new Q-tensor model and the numerical methods to compute the energy functional. In section \ref{section: phy}, the numerical results regarding $\pI$-$\pN$-$\pA$-$\pC$ phase transitions and $\pI$-$\pN$ interface are presented. In section \ref{section: compare}, we compare the Q-tensor model with the classical models. We give several concluding remarks in section \ref{section: sum}. Some detailed calculations involved in the paper are provided in the appendix. 

\subsection{Notations}
In this paper, all vectors will be expressed by boldface letters. The juxtaposition of a pair of vectors $\tm\ttn$ denotes the dyadic product of $\tm$ and $\ttn$. Doubly contracted tensor products are represented by a colon. The Einstein summation convention for tensors is used.

\section{Model}\label{section: prop}

In this section, we review the new Q-tensor model previously derived from Onsager's molecular theory in the paper \cite{HLWZ}, and discuss the numerical methods. We will demonstrate the merits of density as the order parameter in terms of concise mathematics, efficient computations, and explanatory physics. 

\subsection{Mathematical Properties} \label{section: math}

The new Q-tensor model gives the free energy of liquid crystals as equation \refequ{energy0}, with the definitions of notations as follows. Consider the rod-like liquid crystal molecules in domain $\Omega \subset {\mathbb{R}}^3$.  The molecules take positional coordinate $\tx \in \Omega$, and orientational coordinate $\tm \in \mathbb{S}^2$. The positional coordinate $\tx = \bar \tx/L$ is dimensionless, which is the ratio of the physical position $\bar \tx$ and the length of the molecules $L$. We introduce the density $c(\tx)$ and the Q-tensor $Q(\tx)$ as order parameters. The density $c(\tx)$ is dimensionless, representing the number of molecules in volume $L^3$. The tensor $Q(\tx)$ is the second moment of $\tm$ with respect to $\rho$,
\begin{equation}\label{defQ}
Q(\tx) = \int_{\mathbb{S}^2} (\tm \tm -\frac{1}{3}\mathbf{I}) \rho(\tx, \tm) \ud \tm,
\end{equation}
where $\rho$ is the probability density function of molecules with respect to orientation $\tm$ at given position $\tx$. This definition requires the Q-tensor to be a symmetric, traceless $3 \times 3$ matrix whose eigenvalues $\{ \lambda_i (Q)\}$ are constrained by the inequalities
\begin{equation}\label{Qinequ}
-\frac{1}{3} < \lambda_i(Q) < \frac{2}{3}, \quad i=1,2,3; \quad \sum_{i=1}^3 \lambda_i(Q)=0.
\end{equation}

Let $Q_4$ be the fourth moment of $\tm$ in terms of $\rho$, which is defined as
\begin{equation}\label{defQ4}
Q_{4ijkl}(\tx) = \int_{\mathbb{S}^2} (m_i m_j m_k m_l -\frac{1}{7}(m_i m_j \delta_{kl})_{sym} + \frac{1}{35} (\delta_{ij}\delta_{kl})_{sym}) \rho(\tx, \tm) \ud \tm,
\end{equation}
where $(\cdot)_{sym}$ denotes the symmetrization of the tensor with respect to all permutations of indices.

Various closure models were proposed to represent the relationship between $Q_4$ and $Q$  \cite{closure}. Among various closure models, we choose the Bingham closure model, which has the following good properties.
\begin{itemize}
\item The maximizer $\rho^*(\tx, \tm)$, maximizing the entropy $\int_{\mathbb{S}^2} \rho \ln \rho \ud \tm$ subject to the Q-tensor $Q(\tx)$ conforming \refequ{defQ} and \refequ{Qinequ}, follows the Bingham distribution \cite{paper: jball}. 
\item The Bingham closure automatically ensures \refequ{Qinequ}, which preserve the physical meanings of the Q-tensor. 
\item The dynamic Q-tensor model, derived from Doi's kinetic theory using the Bingham closure, obeys the energy dissipation law \cite{HLWZ}.
\item Some numerical results show that the Bingham closure gives the best approximation to Doi's kinetic theory in simulating complex flows of liquid-crystalline polymers \cite{closure}
\end{itemize}

The Bingham distribution of $\rho(\tx, \tm)$ reads
\begin{equation}\label{defrho}
\rho(\tx, \tm) = \frac{1}{Z(\tx)} \exp(B_Q(\tx) : \tm \tm),
\end{equation}
where
\begin{equation}
Z(\tx) = \int_{\mathbb{S}^2} \exp(B_Q(\tx) : \tm \tm) \ud \tm,
\end{equation}
and $B_Q(\tx)$ is any traceless symmetric matrix in $\mathbb R^{3 \times 3}$. Under this assumption, it has been proven that given any $Q(\tx)$ which conforms \refequ{Qinequ}, $B_Q(\tx)$ and $\rho(\tx, \tm)$ can be uniquely determined \cite{Bingham Prove}. Therefore, the dimension of the probability density function $\rho(\tx, \tm)$ reduces to five, the same as that of $Q(\tx)$ and $B_Q(\tx)$.

As a generalization of the Landau-de Gennes theory, the introduction of the density $c(\tx)$ as another order parameter enables the model to characterize smectic phases and density variations in physical phenomena such as defects and interfaces. The major phases of rod-like molecule liquid crystals can be characterized by $c(\tx)$ and $Q(\tx)$ as
\begin{itemize}
\item $\pI$ phase: $c=\text{const}$, $Q=0$.
\item $\pN$ phase: $c=\text{const}$, $Q=s(\ttn\ttn-\frac{1}{3}\mathbf{I})=\text{const}$.
\item $\pA$ phase: $c$ and $Q$ are one-dimensional and periodic. The director $\ttn$, defined as the principal eigenvector of Q-tensor, is parallel to the layer normal $\nabla c$. 
\item $\pC$ phase: $c$ and $Q$ are one-dimensional and periodic. The director $\ttn$ and the layer normal $\nabla c$ yield angle $\theta$, which represents the average of tilt angles between the rod-like molecules and the layer normal. 
\end{itemize}
Biaxiality is an important property in smectic phases, and in other physical phenomena such as defects and interfaces. In this paper, biaxiality is given by a common mathematical definition
\begin{equation}\label{Biaxiality}
B(\tx)=1- 6\frac{(\tr(Q(\tx)^3))^2}{(\tr(Q(\tx)^2))^3}.
\end{equation}
This definition requires the biaxiality $B(\tx)$ within the interval $[0,1]$. If the Q-tensor is uniaxial, $B(\tx)$ will be zero; if the Q-tensor shows strong biaxiality, $B(\tx)$ will tend to one.

Other notations in the energy functional \refequ{energy0} are as follows. $\beta=\frac{1}{k_BT}$ is the thermodynamic beta. The $14$ dimensionless coefficients $A_i$, $G_i$, and $H_i$ are determined in terms of the microscopic molecule's interaction potential, and are functions of temperature. We can infer their approximate range of values by assuming the interaction potential to be the volume exclusion potential of rigid rod-like molecules \cite{HLWZ}. To ensure that the energy is bounded from below, it is necessary that $H_1$ and $H_2 + \frac{2}{3}H_3+\frac{2}{3}H_4$ be positive. Otherwise, assuming the order parameters to be a series of highly oscillating functions, it is easy to show that the energy functional tends to negative infinity.

\subsection{Numerical Methods} \label{section: comp}

We consider the one-dimensional Q-tensor model. We note here that the one-dimensional case is significant and sufficiently representative, since $\pI$, $\pN$, $\pA$, and $\pC$ phases can all be represented in one dimension. For periodic boundary conditions, the spectral method is efficient and accurate enough to compute the energy functional, by representing the order parameters with a few Fourier modes. As another method, we can discretize the order parameters in nodal space and use finite difference method to calculate the derivatives. 

Consider the reduced energy functional on the interval $[0, h]$, which takes the form
\begin{equation}\label{equation reduced}
\begin{aligned}
\beta F_h[c,Q]&= \int_0^h c(x)(\ln c(x) + B_Q:Q-\ln Z(x)) \ud x\\
&+\frac{1}{2} \int_0^h \Big[A_{1}c^2 -A_{2} ( cQ_{ij})^2-A_{3}( cQ_{4ijkl})^2\Big] \ud x \\
&+\frac{1}{2}\int_0^h \Big[ -G_1 ( \frac{\ud}{\ud x} c)^2 +G_2 ( \frac{\ud}{\ud x}(cQ_{ij}))^2+G_3(\frac{\ud}{\ud x}(cQ_{1k}))^2 -G_4 \frac{\ud}{\ud x}(cQ_{11})\frac{\ud}{\ud x}(c)\\
&\quad\quad+G_5( \frac{\ud}{\ud x}(cQ_{4ijkl}))^2 +G_6(\frac{\ud}{\ud x}(cQ_{41klm}))^2+G_7 \frac{\ud}{\ud x}(cQ_{411kl})\frac{\ud}{\ud x}(cQ_{kl}) \Big] \ud x\\
&+\frac{1}{2}\int_0^h \Big[ H_1(\frac{\ud^2}{\ud x^2}(c))^2+H_2(\frac{\ud^2}{\ud x^2}(cQ_{pq}))^2+H_3(\frac{\ud^2}{\ud x^2}(cQ_{11}))^2 +H_4(\frac{\ud^2}{\ud x^2}(cQ_{1p}))^2 \Big] \ud x .
\end{aligned}
\end{equation}

We consider how to numerically represent the order parameters. In the following, we will show that $Q(x)$ can be represented by five independent scalar variables $\lambda(x)$, $d(x)$, $\theta(x)$, $\beta(x)$, and $\gamma(x)$.

The Bingham assumption implies that $B_Q$, $Q$, and $Q_4$ are mutually determined and can be diagonalized simultaneously. Let the eigenvalue decomposition of $B_Q$ be $B_{Q , mn} = \hat B_{ij} T_{im} T_{jn}$, where $\hat B$ is a traceless diagonal matrix and $T$ is an orthogonal matrix. Using a change of variables in the integral \refequ{defQ} and \refequ{defQ4}, $Q$ and $Q_4$ can be represented as
\begin{equation}
\begin{aligned}
Q_{mn}&= \hat Q_{ij} T_{im} T_{jn},\\
Q_{4mnop}&=\hat Q_{4ijkl} T_{im} T_{jn} T_{ko} T_{lp}.\\
\end{aligned}
\end{equation}
Here, $\hat Q$ and $\hat Q_4$ are the diagonal forms of $Q$ and $Q_4$ respectively, which can be computed using $\hat B$ as
\begin{equation}
\begin{aligned}
\hat Q_{ij} &=\frac{1}{Z}\int_{\mathbb{S}^2}(m_i m_j-\frac{1}{3}\delta_{ij})\exp(\hat B:\tm\tm) \ud\tm,\\
\hat Q_{4ijkl}&=\frac{1}{Z}\int_{\mathbb{S}^2}\Big(m_im_jm_km_l-\frac{1}{7}(m_im_j\delta_{kl})_{sym}+\frac{1}{35}(\delta_{ij}\delta_{kl})_{sym}\Big)\exp(\hat B:\tm\tm) \ud \tm,\\
Z &= \int_{\mathbb{S}^2}\exp(\hat B:\tm \tm )\ud \tm.
\end{aligned}
\end{equation}

We use $\lambda$ and $d$ to represent the traceless diagonal matrix $\hat B$, 
\begin{equation}
\hat B = \mathrm{diag}(\lambda, -\frac{1}{2}\lambda+d, -\frac{1}{2}\lambda-d),
\end{equation}
and use Euler angles $\theta$, $\beta$, and $\gamma$ to represent the orthogonal transformation matrix $T$,
\begin{equation}
T=\left(
    \begin{array}{ccc}
      \cos\theta & -\sin\theta\cos\gamma & \sin\theta\sin\gamma  \\
      \sin\theta\cos\beta & \cos\theta\cos\beta\cos\gamma-\sin\beta\sin\gamma & -\cos\theta\cos\beta\sin\gamma-\sin\beta\cos\gamma \\
      \sin\theta\sin\beta & \cos\theta\sin\beta\cos\gamma+\cos\beta\sin\gamma & -\cos\theta\sin\beta\sin\gamma+\cos\beta\cos\gamma \\
      \end{array}
  \right).
\end{equation}
Here, the first angle $\theta$ is the angle between the principal eigenvector of $Q$ and the x-axis. 

Above all, the Q-tensor $Q(x)$ can be represented by $\lambda(x)$, $d(x)$, $\theta(x)$, $\beta(x)$, and $\gamma(x)$. In the meanwhile, $B_Q(x)$, $Z(x)$, and $Q_4(x)$ can also be computed using these five variables.

One way to discretize the order parameters is to expand them by Fourier modes
\begin{equation}
\begin{aligned}
c(x) &= c_0+\sum_{j=1}^{n} c_j^{(1)} \cos jkx + c_j^{(2)} \sin jkx,\quad
\lambda(x) = \lambda_0+\sum_{j=1}^{n} \lambda_j^{(1)} \cos jkx+\lambda_j^{(2)} \sin jkx,\\
d(x) &= d_0+\sum_{j=1}^{n} d_j^{(1)} \cos jkx+d_j^{(2)} \sin jkx,\quad
\theta(x) = \theta_0+\sum_{j=1}^{n} \theta_j^{(1)} \cos jkx+\theta_j^{(2)} \sin jkx,\\
\beta(x) &= \beta_0+\sum_{j=1}^{n} \beta_j^{(1)} \cos jkx+\beta_j^{(2)} \sin jkx,\quad
\gamma(x) = \gamma_0+\sum_{j=1}^{n} \gamma_j^{(1)} \cos jkx+\gamma_j^{(2)} \sin jkx,\\
\end{aligned}
\end{equation}
where $k = 2\pi/h$. The dimension of variables is $12n + 6$. The derivatives of $c \cdot Q$ and $c\cdot Q_4$ are calculated using Fast Fourier transform (FFT). In the following section, we use this method in computing the phase transition problem; empirically, $n=6$ is enough to characterize the profiles of the order parameters. 

As another way to represent the order parameters, we can discretize the interval $[0, h]$ with $N$ nodes. The dimension of variables is $6N$. The derivatives of the order parameters are computed using the finite difference method. In the following section, we use this method in computing the $\pI$-$\pN$ interface problem, and $N=50$ is accurate enough empirically.

The minimum of the energy functional can be found by standard methods. We use numerical differentiation to calculate the gradient of the objective function, and use quasi-Newton methods, such as the BFGS method to solve the optimization problem. The computations for both of the two problems discussed in the following section converge in about 100 steps with absolute precision of $1\times 10^{-5}$. The overall computational cost is as moderate as the cost of the Landau-de Gennes model. 

It is worth pointing out that the Q-tensor model may have some local minima. In the simulation, we try different random initial values to ensure that we find the global minimum. In practice, there are no other local minimum points near the global minimum point.

\section{Results and Discussions} \label{section: phy}

In the paper \cite{HLWZ} where the new Q-tensor model was first introduced, the coefficients $A_i$, $G_i$, and $H_i$ were expressed in terms of the molecule's shape factor $\varepsilon = D/L$, under the assumption that the interaction is the hard core potential of the rod-like molecules without attraction effects, with $D$ the diameter of semi-sphere at two ends of the rod and $L$ the length of the rod. If we take account of the attraction effects between molecules, these coefficients would be functions of temperature, and alter within a few order of magnitudes as the temperature changes \cite{paper:symmetry}. 

In this paper, we regard the new Q-tensor model as a phenomenological model. We set these coefficients near the values deduced using the hard core potential. Fix
\begin{equation}\label{coefficients}
\begin{aligned}
&A_1=0.22409, \quad A_2= 0.14728, \quad A_3= 0.09663,\\
&G_1=0.00884, \quad G_2=0.00182, \quad G_3= 0.00680,\\
&G_5=0.00016, \quad G_6=0.00174, \quad G_7=0.01181.\\
&H_1= 0.00028, \quad H_2=H_4=0.\\
\end{aligned}
\end{equation}
The coefficients $A_i$ and $G_i$ are derived in terms of microscopic shape factor $\varepsilon=D/L = 0.1$. We set $H_1= 0.00028$ to ensure the lower boundedness, and set $H_2=0$ and $H_4=0$ since they are order of magnitude smaller. Two remaining coefficients $G_4$ and $H_3$ are not fixed. Empirically, they are more important to determine the phases of liquid crystals. We will set these two coefficients near $G_4^{(0)}=0.02374$ and $H_3^{(0)}=0.00063$ within an order of magnitude. Note that we cannot relate the coefficients $A_i$, $G_i$, $H_i$ with any specific liquid crystal materials so far, and the relationship is still under investigation. 

We consider the phase transition problem and the $\pI$-$\pN$ interface problem. The phase transition problem is to find the most stable phase of liquid crystals given the average density, to find the characterizations of the various phases, and to study the transition of phases from one to another as the average density increases. In some suitable environments, $\pI$ and $\pN$ phases will coexist and take up different regions in the liquid crystal materials. The $\pI$-$\pN$ interface problem is to investigate the interfacial region between the $\pI$ and $\pN$ phases. The mathematical formulation of these two problems will be given below. Density variations are essential in both problems. In the phase transition problem, the density variation serves to lower the second order elastic energy $F_{elastic, 2}$, which renders the smectic phase stable. In the $\pI$-$\pN$ interface problem, given the chemical potential and the grand potential density, the density at $\pI$ and $\pN$ phases are different, which renders a natural variation of density at the interface.

\subsection{Phase Transitions}

First, we consider the phase transition problem. Since $\pI$ and $\pN$ phases are spatially homogeneous, and $\pA$ and $\pC$ phases are layered, we presume the solution to be periodic to reduce the computational costs. Therefore, we consider the domain $\Omega=[0,h]$, and assume that the order parameters enjoy the periodic boundary conditions. Given the average density $c_0$, we need to minimize the average free energy $F_h / h$, where $F_h$ is defined as equation \refequ{equation reduced}. 

We have the optimization problem 
\begin{equation}
\begin{aligned}
\min_{c(x),Q(x),h}\quad \qquad &\left \{ \frac{F_h}{h} \right \},\\
\text{subject to}\quad \qquad  c_0&=\frac{1}{h}\int_0^h c(x) \ud x, \\
c(0) &= c(h)\\
Q(0) &= Q(h).\\
\end{aligned}
\end{equation}

\begin{figure}
\centering
\includegraphics[width=0.8\linewidth]{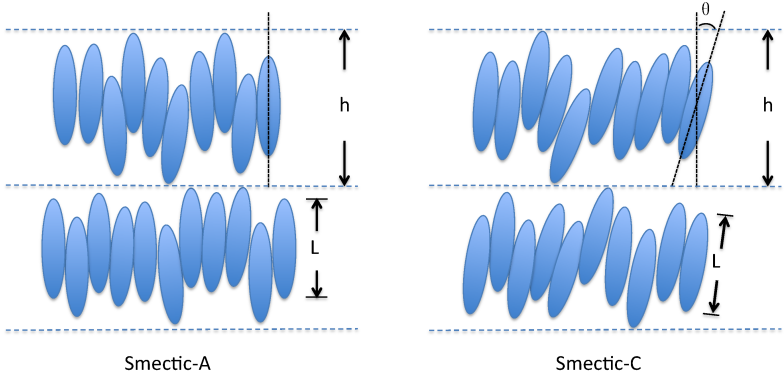}
\caption{Illustrations for smectic phases. The tilt angle $\theta$ and the layer thickness $h$ are their important characterizations.}\label{illustration smectic}
\end{figure}

Spectral methods are used to calculate the free energy. The period $h$, which is also a variable that needs to be optimized, represents the layer thickness in the smectic phases. \reffig{illustration smectic} illustrates the microscopic configuration of molecules of the smectic phases. The following phase transitions are found given various sets of $G_4$ and $H_3$.
\begin{itemize}
\item $G_4=G_4^{(0)}\times 3.9$, $H_3= H_3^{(0)} \times 1.7$.
\begin{center} 
$\pI \xlongleftrightarrow{c \approx 45.2} \pN \xlongleftrightarrow{c \approx 46.5} \pA \xlongleftrightarrow{c \approx 50.3} \pC$.
\end{center}
\item $G_4=G_4^{(0)}$, $H_3= H_3^{(0)}$.
\begin{center} 
$\pI \xlongleftrightarrow{c \approx 45.2} \pN \xlongleftrightarrow{c \approx 50.6} \pA$.
\end{center}
\item $G_4=G_4^{(0)}\times 2$, $H_3= H_3^{(0)} \times 6$.
\begin{center} 
$\pI \xlongleftrightarrow{c \approx 45.2} \pN \xlongleftrightarrow{c \approx 50.6} \pC$.
\end{center}
\end{itemize}

\begin{figure}
\begin{minipage}{0.48\linewidth}
\begin{center}
\includegraphics[width=\linewidth]{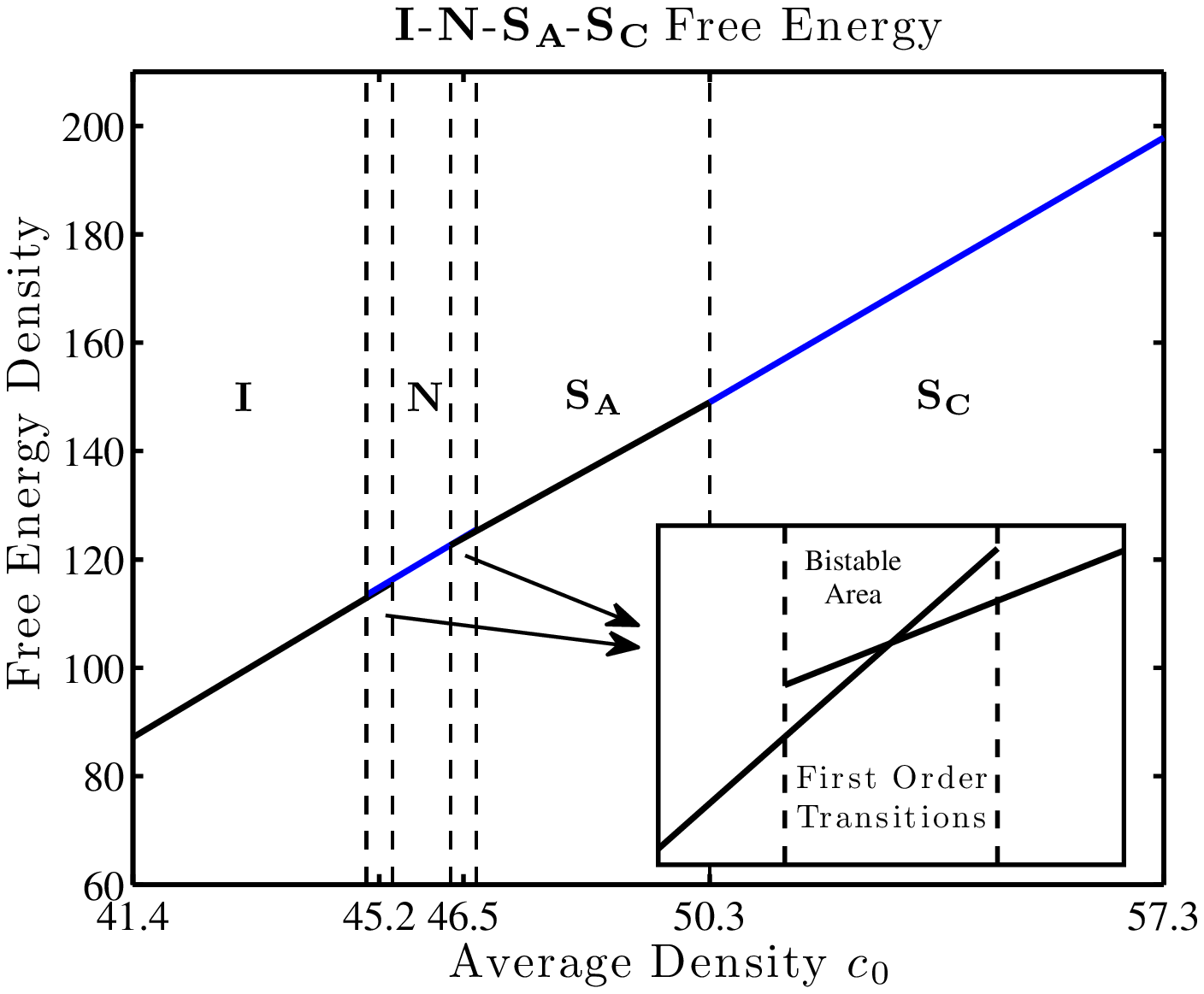}
\end{center}
\caption{An one-dimensional phase diagram with four phases presented. The small regions between the dashed line represents the bistable areas. The remaining coefficients are $G_4=G_4^{(0)}\times 3.9$ and $H_3= H_3^{(0)} \times 1.7$.}\label{fig:energy vs density}
\end{minipage}
\begin{minipage}{0.48\linewidth}
\begin{center}
\includegraphics[width=\linewidth]{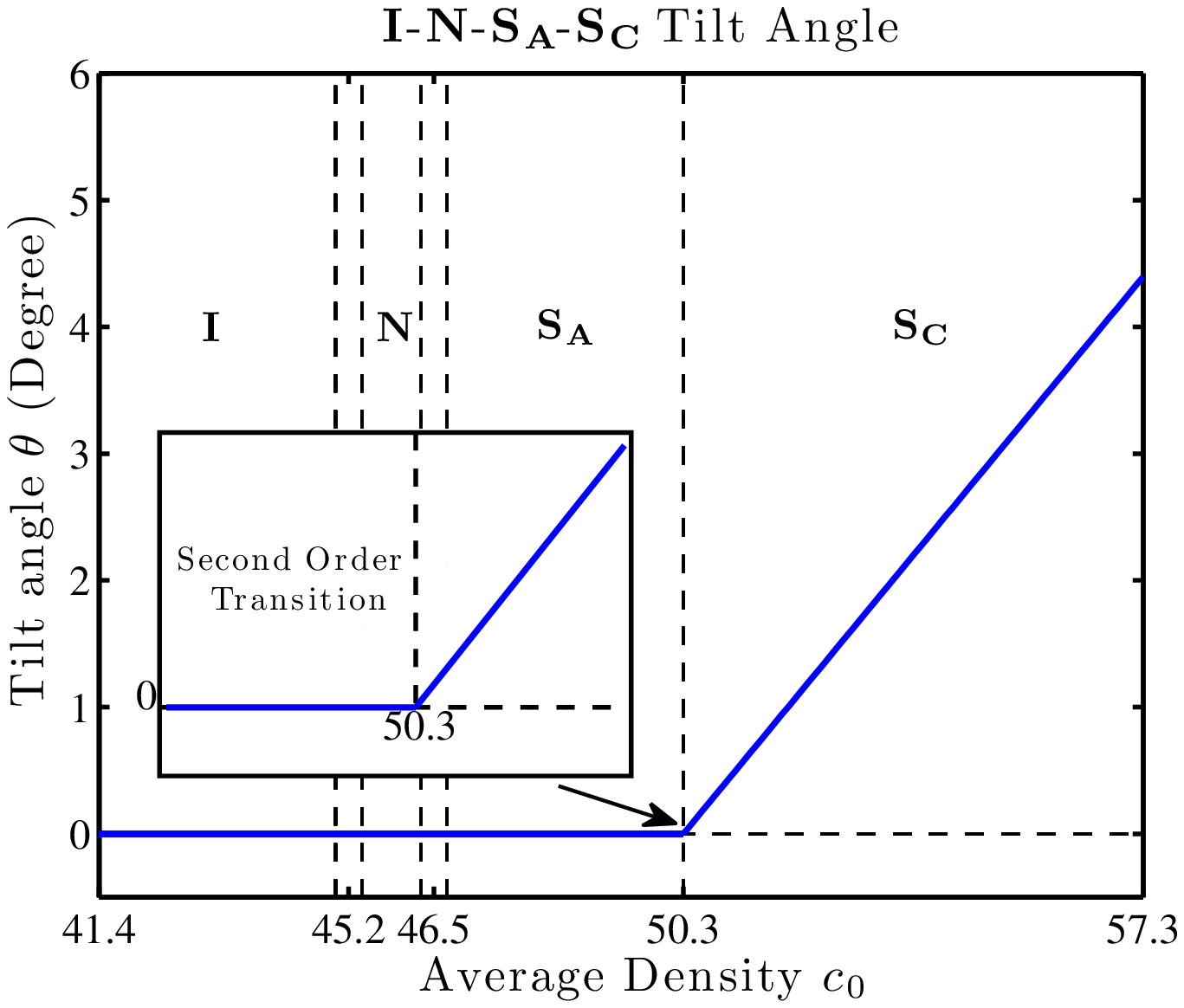}
\caption{The tilt angle versus average density. The tilt angle of smectic phase increases continuously after the average density exceeds 50.3. The $\pA$-$\pC$ transition is second order.}\label{fig:angle vs density}
\end{center}
\end{minipage}
\end{figure}

The one-dimensional phase diagrams of $\pI$-$\pN$-$\pA$-$\pC$ transitions are presented using $F_{min} - c_0$ coordinates in \reffig{fig:energy vs density}, and using $\theta - c_0$ coordinates in \reffig{fig:angle vs density}.

As the average density $c_0$ increases, the system exhibits $\pI$, $\pN$, $\pA$, and $\pC$ phases sequentially. As shown in \reffig{fig:energy vs density}, $\pI$-$\pN$ and $\pN$-$\pA$ phase transitions are first order. $\pI$ and $\pN$ phases coexist as $c_0$ is within $45.2 \pm 0.2$; $\pN$ and $\pA$ phases coexist as $c_0$ is within $46.5 \pm 0.2$. The free energies of both phases in these regions are local minima, which are confirmed by numerically evaluating the Hessian matrix at the minimum points. It is noteworthy that the $\pN$-$\pA$ phase transition is of weak first order: at both of the local minimum points, the smallest eigenvalue of the Hessian is near 0, which indicates that the energy barrier between these two stable phases is rather small. From the physical point of view, the average density $c_0 = 45.2$ states that on average there are $45.2$ liquid crystal molecules with shape factor $\varepsilon = 0.1$ in a cube with volume $L^3$, and the molecules take up roughly $35$ percent the volume of the entire space. 

As shown in \reffig{fig:angle vs density}, the $\pA$-$\pC$ phase transition is of second order. After $c_0$ exceeds $50.3$, the tilt angle of director increases from $0$ to $5$ degrees continuously. In one layer, the centers of molecules align in order, but not exactly in a plane, which renders the layer thickness larger than the length of molecules. The optimized layer thickness for both $\pA$ and $\pC$ phases is roughly $1.81L$, i.e. $1.81$ times the length of the molecule, and it varies slightly within the order $0.005L$ as the average density increases. The layer thickness is quite sensitive to the coefficients $G_i$ and $H_i$. If $H_3$ decreases from $H_3^{(0)} \times 3.9$ to $H_3^{(0)} \times 3.6$, the layer thickness will decrease from $1.81L$ to $1.74L$. This relation between the coefficients and the layer thickness is explained using the vector model, with its details discussed in Section \ref{section: CL}.

\begin{figure}
\begin{center}
\includegraphics[width=0.6\linewidth]{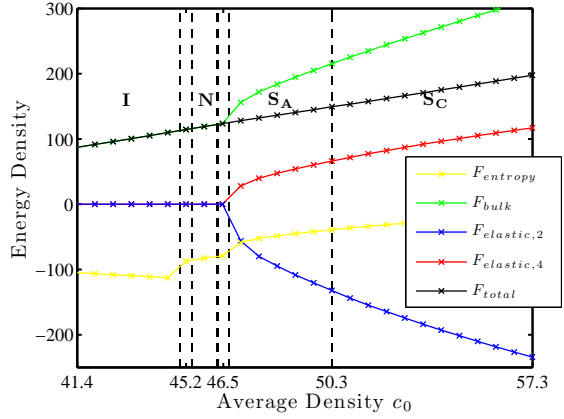}
\end{center}
\caption{The components of free energy and the total free energy.}\label{separ}
\end{figure}

\reffig{separ} shows the components of the free energy as the average density increases, which exhibits the order of transitions more clearly. The definitions of these components are given as \refequ{bulk}, \refequ{elastic2}, and \refequ{elastic4}, and the entropy is the first part of the bulk energy $F_{bulk}$. The entropy decreases at the beginning. It has positive leaps at $\pI$-$\pN$ and $\pN$-$\pA$ critical points, and finally increases gradually. The bulk energy increases gradually with a leap at $\pN$-$\pA$ critical point. The elastic energy $F_{elastic, 2}$ and $F_{elatic, 4}$ are zero in $\pI$ and $\pN$ phases. The second order elastic energy $F_{elastic, 2}$ drops and continuously decreases after reaching $\pN$-$\pA$ critical point, while the fourth order elastic energy $F_{elastic, 4}$ soars up and continuously increases after $\pN$-$\pA$ point, with a smaller magnitude than $F_{elastic, 2}$. All these components at $\pA$-$\pC$ critical point are continuous. The total energy increases steadily without significant leaps. 

The second order elastic energy $F_{elastic, 2}$ contains two parts: derivative terms in the density $c(x)$ and the Oseen-Frank energy which contains the derivative terms of director $\ttn$. This decomposition is further discussed in Section \ref{section: CL}. In smectic phases, the Oseen-Frank energy is zero for there is no distortion of the director. The derivative terms in the density $c(x)$ in $F_{elastic, 2}$ with negative coefficients are the reason for the density-varied phases to have a lower free energy. The second order derivative terms in the density in $F_{elastic, 4}$ with positive coefficients are the reason for the total energy to be stable.

\begin{figure}
\begin{minipage}{0.48\linewidth}
\centering
\includegraphics[width=\linewidth]{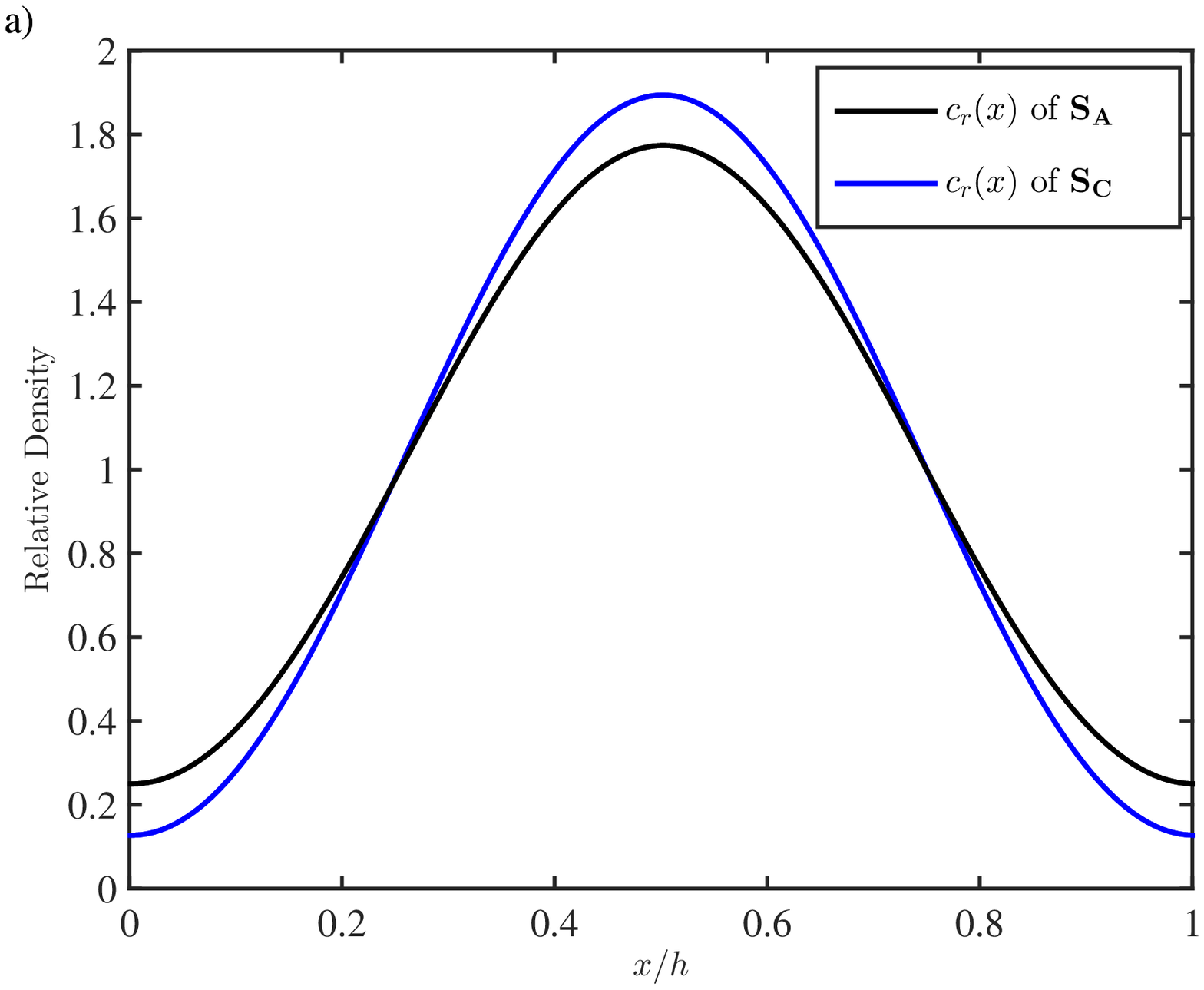}
\end{minipage}
\begin{minipage}{0.48\linewidth}
\centering
 \includegraphics[width=\linewidth]{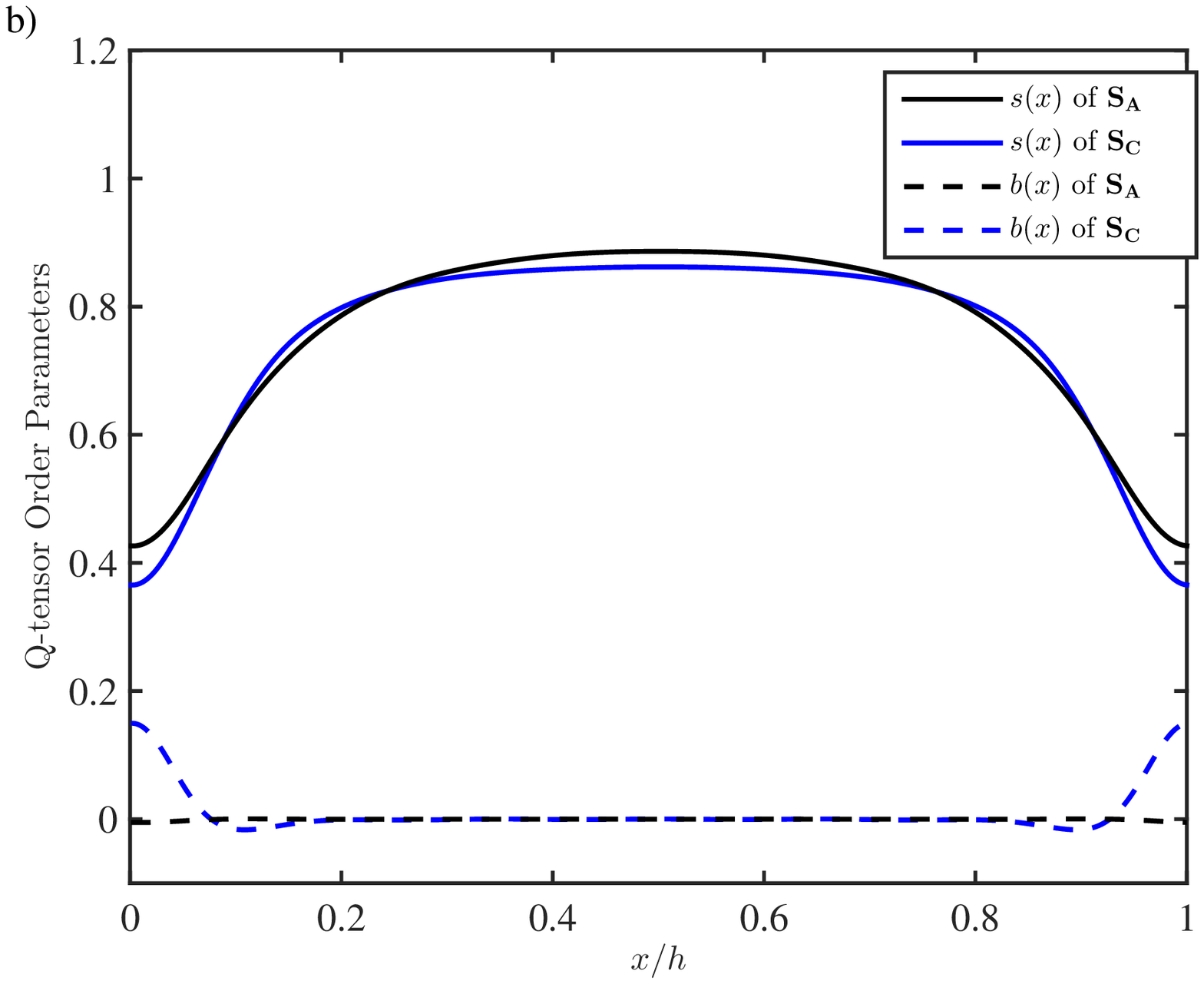}
\end{minipage}
\begin{minipage}{0.48\linewidth}
\centering
 \includegraphics[width=\linewidth]{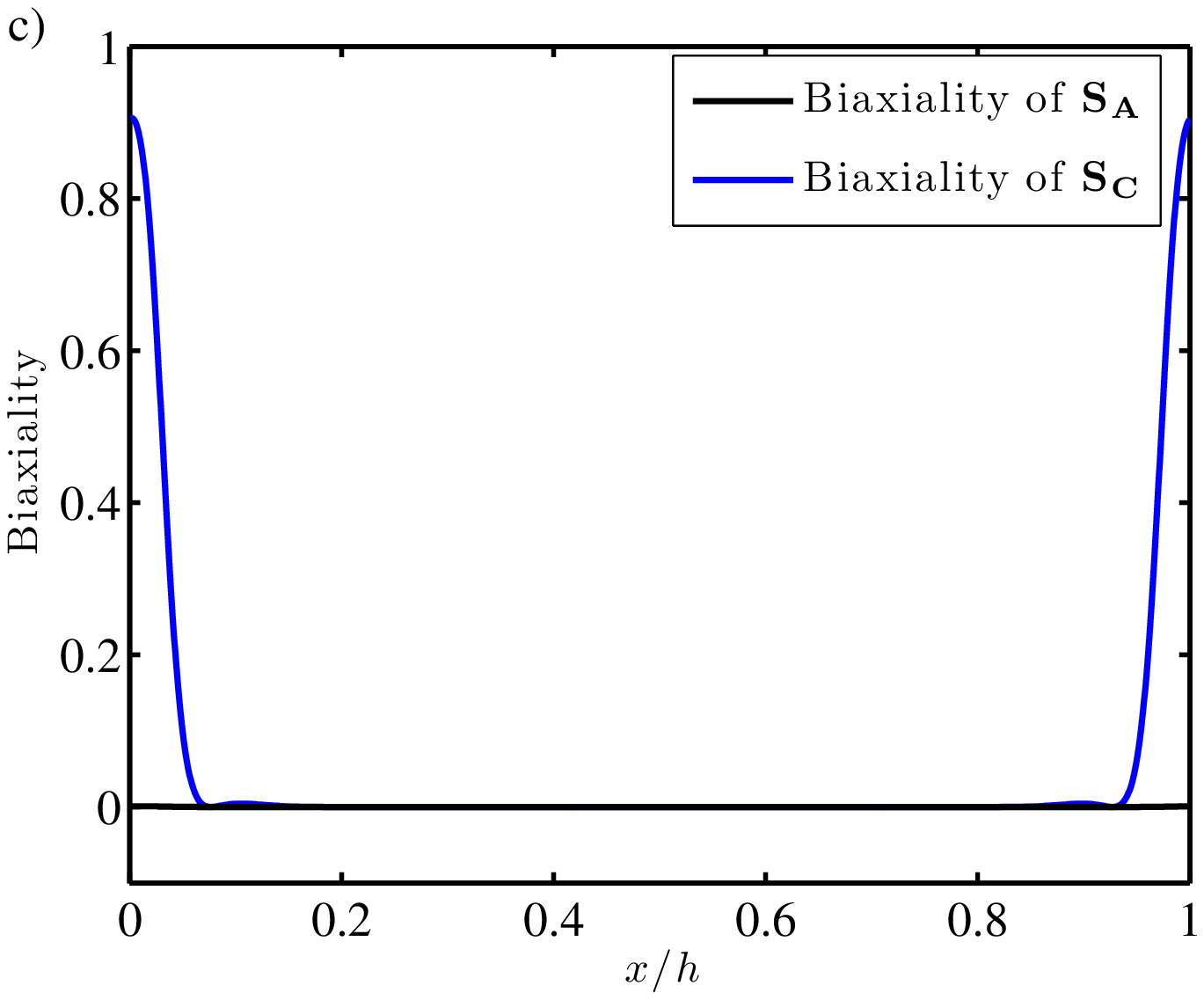}
\end{minipage}
\centering
\caption{The characterizations of the smectic phases, with $x$-axis in the direction normal to the layer.  (a) The relative density $c_r$, defined as $c_r(x)=c(x)/c_0$. (b) The eigenvalues of Q-tensor, written as $Q = s (\mathbf{nn} -\frac{1}{3}\mathbf{I}) + b(\mathbf{n'n'}-\frac{1}{3}\mathbf{I})$, where $\mathbf{n'}$ is perpendicular to $\ttn$; $s$ is the principal nematic order parameter, and $b$ characterizes biaxial effects. (c) The biaxiality, defined as $B(x)=1- 6\frac{(\tr(Q^3))^2}{(\tr(Q^2))^3}$.  }\label{fig:s}
\end{figure}

The representative use of the order parameters and the biaxiality characterizing the smectic phase is shown in \reffig{fig:s}. The relative density $c_r(x)$ is defined as $c_r(x)=c(x)/c_0$. The Q-tensor can be written as $Q = s (\mathbf{nn} -\frac{1}{3}\mathbf{I}) + b(\mathbf{n'n'}-\frac{1}{3}\mathbf{I})$, where $\mathbf{n'}$ is perpendicular to the director $\ttn$; $s(x)$ is the principal nematic order parameter, and $b(x)$ characterizes biaxial effects. The biaxiality $B(x)$ is defined as equation \refequ{Biaxiality}. In the smectic phase, the relative density $c_r(x)$ and order parameter $s(x)$ fluctuate within one layer. The maximum point of $c(x)$ and $s(x)$ are identical, where the centers of molecules concentrate and the molecules are more likely to point to the preferred direction. In the $\pA$ phase, the biaxiality $B(x)$ vanishes, which supports that the uniaxial approximation often adopted in modeling $\pA$. In $\pC$ phase, $B(x)$ vanishes when the relative density $c_r(x)$ is large. At the region between two layers where the relative density $c_r(x)$ is small, the biaxiality $B(x)$ is large. This phenomena can be explained intuitively: the group of molecules at the vicinity region between two layers have more freedom of orientation and are less symmetric than the group of molecules at the center of one layer. 

\begin{figure}
\begin{minipage}{0.48\linewidth}
\centering
\includegraphics[width=\linewidth]{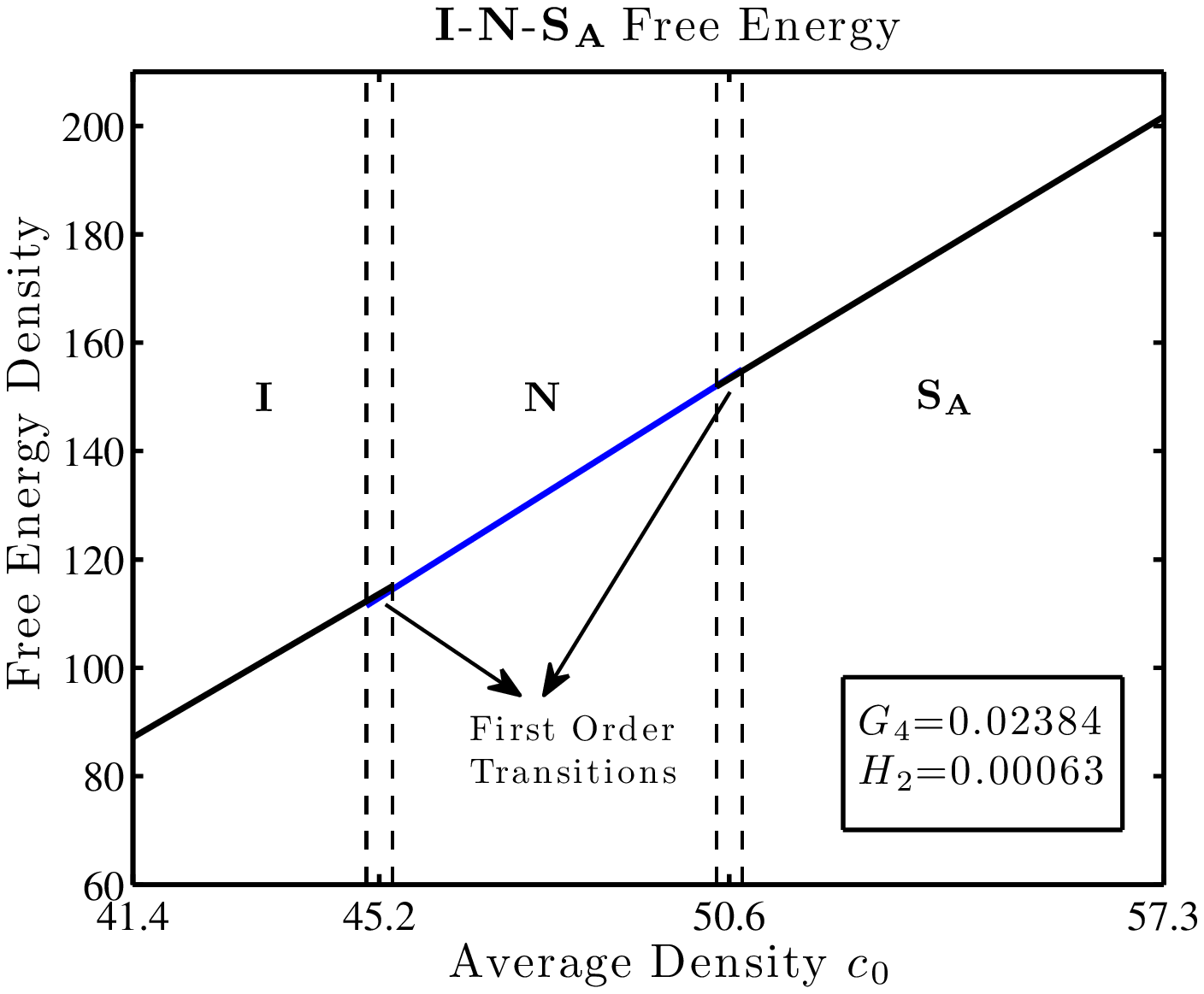}
\caption{An one-dimensional phase diagram of $\pI$-$\pN$-$\pA$  transitions.  The remaining coefficients are $G_4=G_4^{(0)}$ and $H_3= H_3^{(0)}$.}\label{fig:ina}
\end{minipage}
\begin{minipage}{0.48\linewidth}
\centering
 \includegraphics[width=\linewidth]{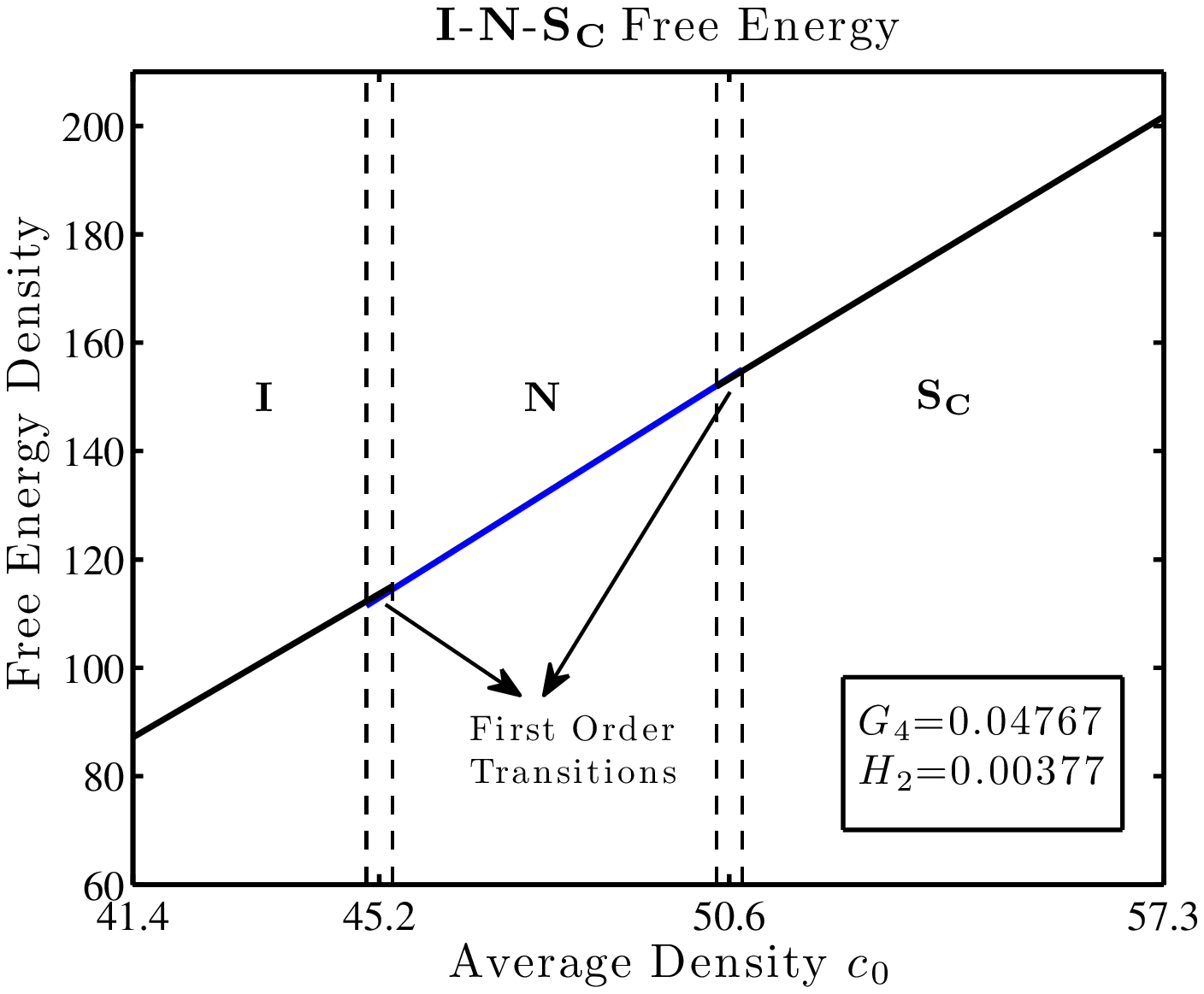}
\caption{An one-dimensional phase diagram of $\pI$-$\pN$-$\pC$ transitions.  The remaining coefficients are $G_4=G_4^{(0)}\times 2$ and $H_3= H_3^{(0)} \times 6$.}\label{fig:inc}
\end{minipage}
\end{figure}

The $\pI$-$\pN$-$\pA$ transitions and $\pI$-$\pN$-$\pC$ transitions are presented given other sets of coefficients. The one-dimensional phase diagrams are presented using $F_{min} - c_0$ coordinates in \reffig{fig:ina} characterizing $\pI$-$\pN$-$\pA$ transitions, and in \reffig{fig:inc} characterizing $\pI$-$\pN$-$\pC$ transitions. In the $\pA$ phase, the layer thickness is roughly $1.52 L$. In $\pC$ phase, the layer thickness is roughly $1.92 L$, and the tilt angle is $23$ degree. The $\pN$-$\pC$ transition is also first order.

According to liquid crystal experiments,  the $\pI$-$\pN$ transition and the $\pN$-$\pA$ transition could be either first or second order, depending on the liquid crystal materials. The $\pI$-$\pN$ transition is most commonly believed to be of first order \cite{book:de gennes}, but some experiments of MBBA [(4-Butyl-phenyl)-(4-methoxy-benzylidene)-amine] reported a second order transitions \cite{INtransition}. The study of 8CB-10CB mixtures \cite{experiment: N-A first order} confirmed the prediction made by Halperin \textit{et al.} that fluctuation change the second order $\pN$-$\pA$ transition into a weak first order one. The study of 4-n-alkoxybenzylidene-4'-phenylazoaniline \cite{experiment:  N-A second order} drew the conclusion that decreasing the length of the alkyl end chain drives the first order $\pN$-$\pA$ transition to be second order. According to Monte-Carlo simulation based on molecular theory, assuming the molecule's shape to be spherocylindrical, McGrother \textit{et al.} \cite{simulation: McGrother}, Bolhuis \textit{et al.} \cite{simulation: Bolhuis}, and Lolson \textit{et al.} \cite{simulation: Lolson} showed that the $\pN$-$\pA$ transition is a first order one, but it tends to become continuous as the shape factor $L/D \rightarrow \infty$. In comparison, our simulation results predict the $\pI$-$\pN$ transition is of first order; the energy barrier between the $\pN$ and $\pA$ phases is rather small, which indicates that the $\pN$-$\pA$ transition is of weak first order.

For the $\pA$-$\pC$ transition, the study of 8S5 [4-n-pentyl-phenylthiol-4'-n-octyloxybenzoate] \cite{experiment: A-C second order} reported a second order transition, and the tilt angle increases from $0$ degree to $20$ degree continuously. The study of C7 [4-(3-methyl-2-chloropentanoyloxy)-4'-heptyloxybiphenyl] \cite{ACfirst} reported a first order $\pA$-$\pC$ transition, and the tilt angle ranges from 0 to 34 degree. In our simulations, the $\pA$-$\pC$ transition is of second order, and the tilt angle ranges from 0 to 23 degree. For the smectic layer thickness, most experiments indicated that the layer thickness of the smectic phase is around the molecule's length $L$. A study of 8CB [4'-n-octyl-4-cyanobiphenyl] and 8OCB [4'-n-octyloxy-4-cyanobiphenyl] \cite{experiment: layer thickness} showed that when molecules form dimers, one smectic layer would contain two molecule layer overlapped a bit, such that the layer thickness of the smectic phase is larger than $L$ but within $2L$. In our simulations, the predicted layer thickness is larger than most experiments results. However, our predicted layer thickness is in a reasonable range, which is larger than the length $L$ of the molecules but within $2L$ varying with different coefficients.

\subsection{$\pI$-$\pN$ Interface}\label{section: interface}

\begin{figure}
\begin{center}
\includegraphics[width=0.55\linewidth]{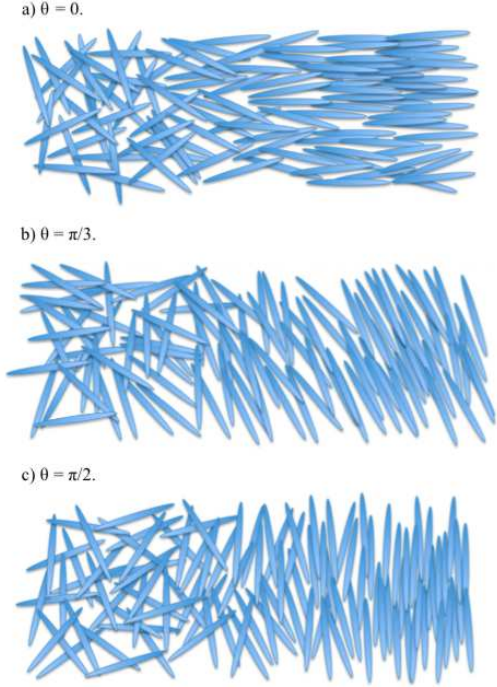}
\end{center}
\caption{Snapshots of configurations in the immediate vicinity of
the $\pI$-$\pN$ interface, for the three different anchoring conditions. The configuration with tilt angle $\theta = \frac{\pi}{2}$ is the most stable.}\label{snapshots}
\end{figure}

Consider the liquid crystals in the entire space. $\pI$ and $\pN$ are at two opposite ends of the space, sharing the same chemical potential and the same grand potential density.  Assume the problem is one dimensional, and the order parameters vary in one dimension and are homogeneous in the other two dimensions. Given the angle $\theta$ between the interface normal and the director of the nematic phase, the configuration of the interface is determined. The microscopic configuration of molecules of $\pI$-$\pN$ interface is illustrated as \reffig{snapshots}. 

Consider the region $\Omega = [-h,h]$. We need to minimize the grand potential 
\begin{equation}\label{grand potential}
\begin{aligned}
\beta \Xi [c, Q]=&\int_{-h}^h c(  x)(\ln c( x) + B_Q:Q-\ln Z( x)) \ud x\\
+&\frac{1}{2}\Big \{ \int_{-h}^h \Big[A_{1}c^2 -A_{2} ( cQ_{ij})^2-A_{3}( cQ_{4ijkl})^2\Big] \ud x \\
+&\int_{-h}^h \Big[ -G_1 ( \frac{\ud}{\ud x} c)^2 +G_2 ( \frac{\ud}{\ud x}(cQ_{ij}))^2+G_3(\frac{\ud}{\ud x}(cQ_{1k}))^2 -G_4 \frac{\ud}{\ud x}(cQ_{11})\frac{\ud}{\ud x}(c)\\
&\quad\quad+G_5( \frac{\ud}{\ud x}(cQ_{4ijkl}))^2 +G_6(\frac{\ud}{\ud x}(cQ_{41klm}))^2+G_7 \frac{\ud}{\ud x}(cQ_{411kl})\frac{\ud}{\ud x}(cQ_{kl}) \Big] \ud x\\
+&\int_{-h}^h \Big[ H_1(\frac{\ud^2}{\ud x^2}(c))^2+H_2(\frac{\ud^2}{\ud x^2}(cQ_{pq}))^2+H_3(\frac{\ud^2}{\ud x^2}(cQ_{11}))^2 +H_4(\frac{\ud^2}{\ud x^2}(cQ_{1p}))^2 \Big] \ud x \Big \}\\
-& \mu \int_{-h}^h c(x) \ud x,
\end{aligned}
\end{equation}
under proper boundary conditions. The chemical potential is chosen to meet the coexistence condition, $\mu = 3.407$. The order parameters for the stable isotropic phase are $c = 43.57$ and $Q=0$; the order parameters for the stable nematic phase are $c = 48.30$ and $Q = s(\ttn \ttn - \frac{1}{3}\pI)$ with $s=0.667$. We anchor the angle $\theta$ between the director and the interface normal to be constant all over the space, and perform simulations with a series of $\theta$.  Thus, the boundary conditions are
\begin{equation}
\begin{array}{c c r c c c r}
c\vert_{x=-h}&=&43.57, &\quad& c\vert_{x=h}&=&48.30, \\
\lambda\vert_{x=-h}&=&0, &\quad&  \lambda\vert_{x=h}&=&3.522, \\
d\vert_{x=-h}&=&0, &\quad& d\vert_{x=h}&=&0 , \\
\beta \vert_{x=-h}&=&0, &\quad& \beta \vert_{x=h}&=&0, \\
\gamma \vert_{x=-h}&=&0,&\quad& \gamma \vert_{x=h}&=&0, \\
\theta(x)&=&\theta_0, & \quad&\\
\end{array}
\end{equation}
where $\theta_0= 0$, $\pi/8$, $\pi/4$, $3\pi/8$, $\pi/2$. The length $h$ is chosen large enough, which is $7L$ in our computations. The coefficients $A_i$, $G_i$, and $H_i$ are given in equations \refequ{coefficients} with $G_4=G_4^{(0)}=0.02374$, $H_3= H_3^{(0)}=0.00063$. Note that the coefficients $G_i$ here are all deduced by assuming the interaction potential to be the volume exclusion potential of rigid rod-like molecules. The results are summarized as follows.

\begin{figure}
\centering
\begin{minipage}{0.48\linewidth}
\begin{center}
\includegraphics[width=\linewidth]{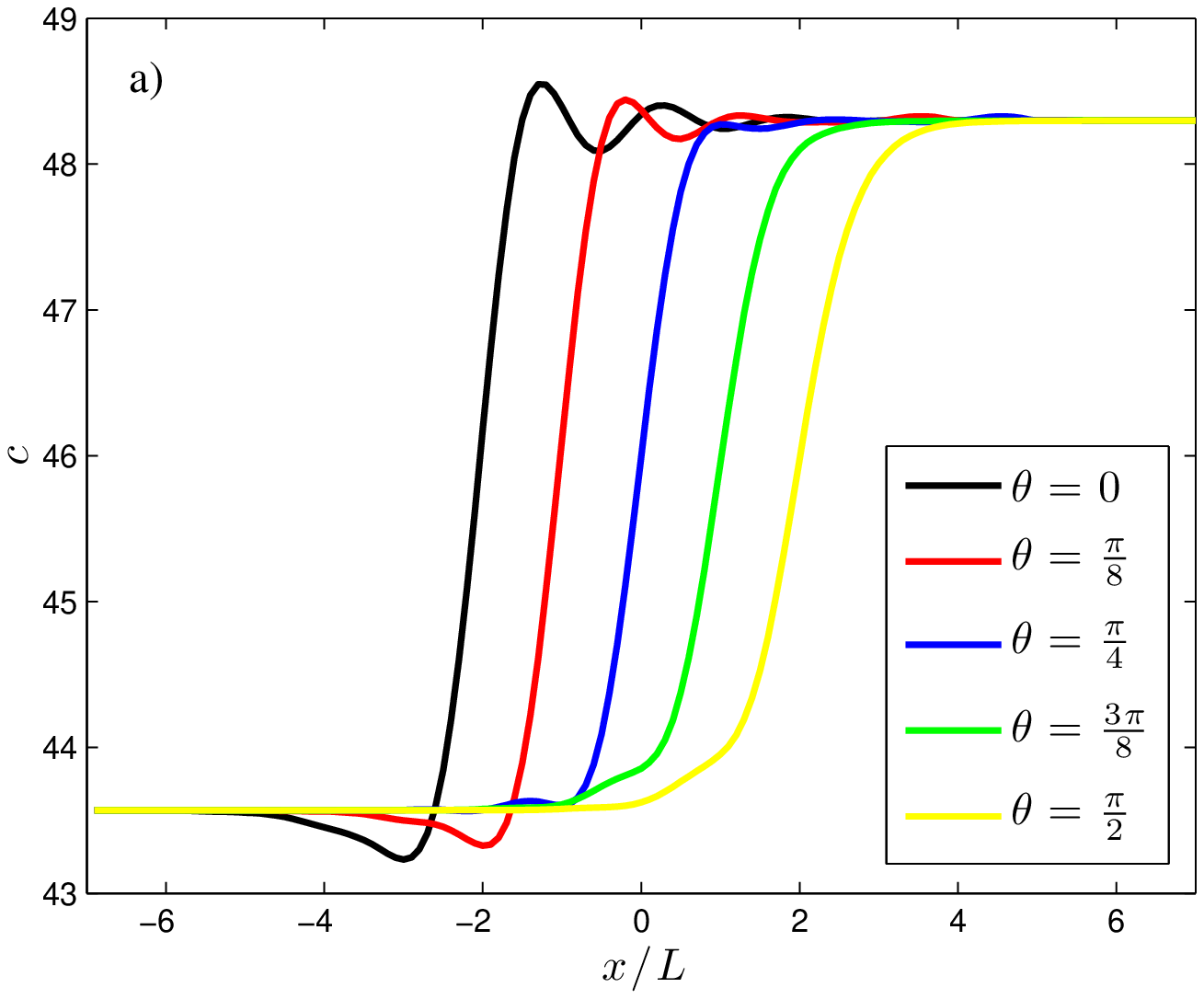}
\end{center}
\end{minipage}
\begin{minipage}{0.48\linewidth}
\begin{center}
 \includegraphics[width=\linewidth]{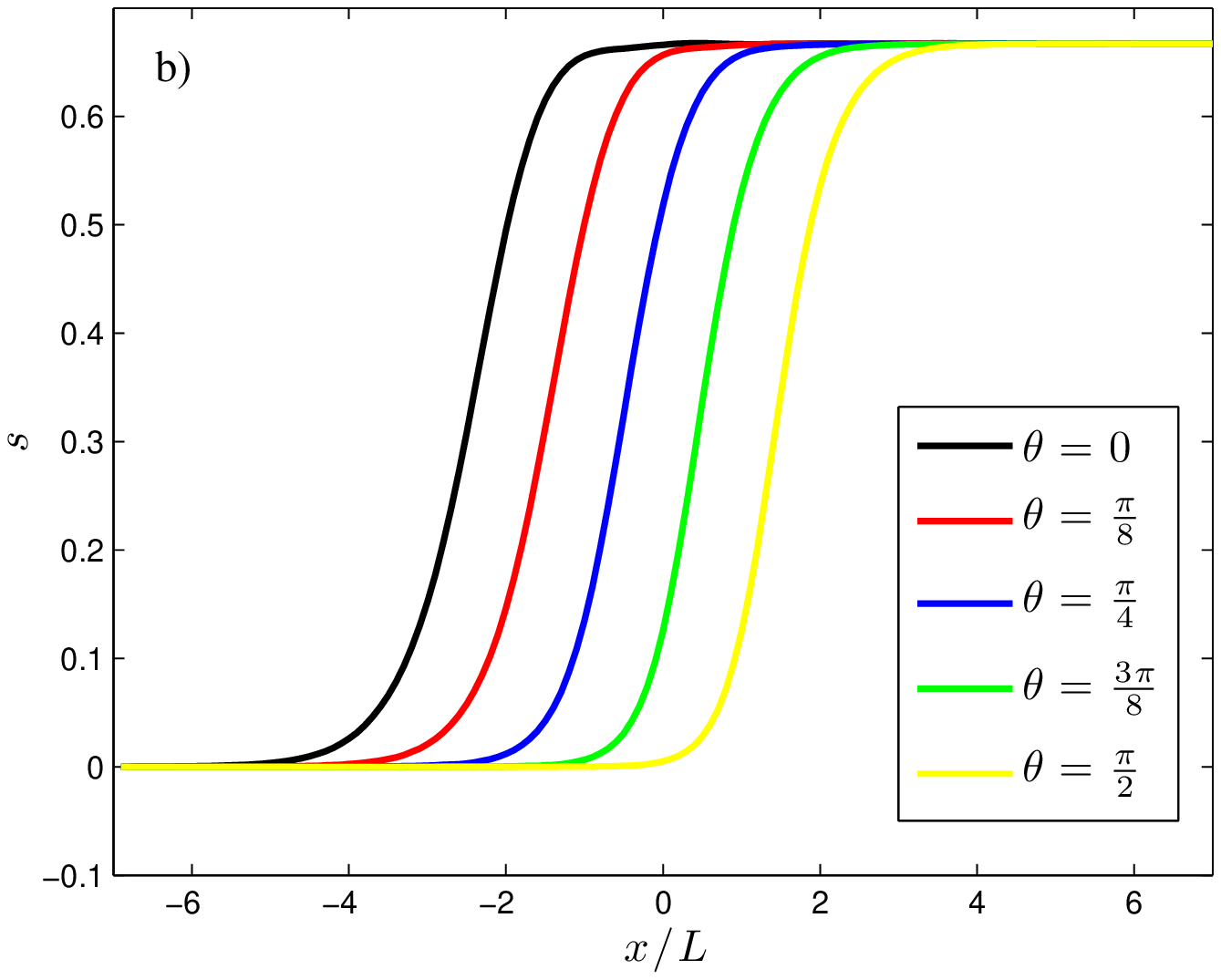}
\end{center}
\end{minipage}
\begin{minipage}{0.48\linewidth}
\begin{center}
 \includegraphics[width=\linewidth]{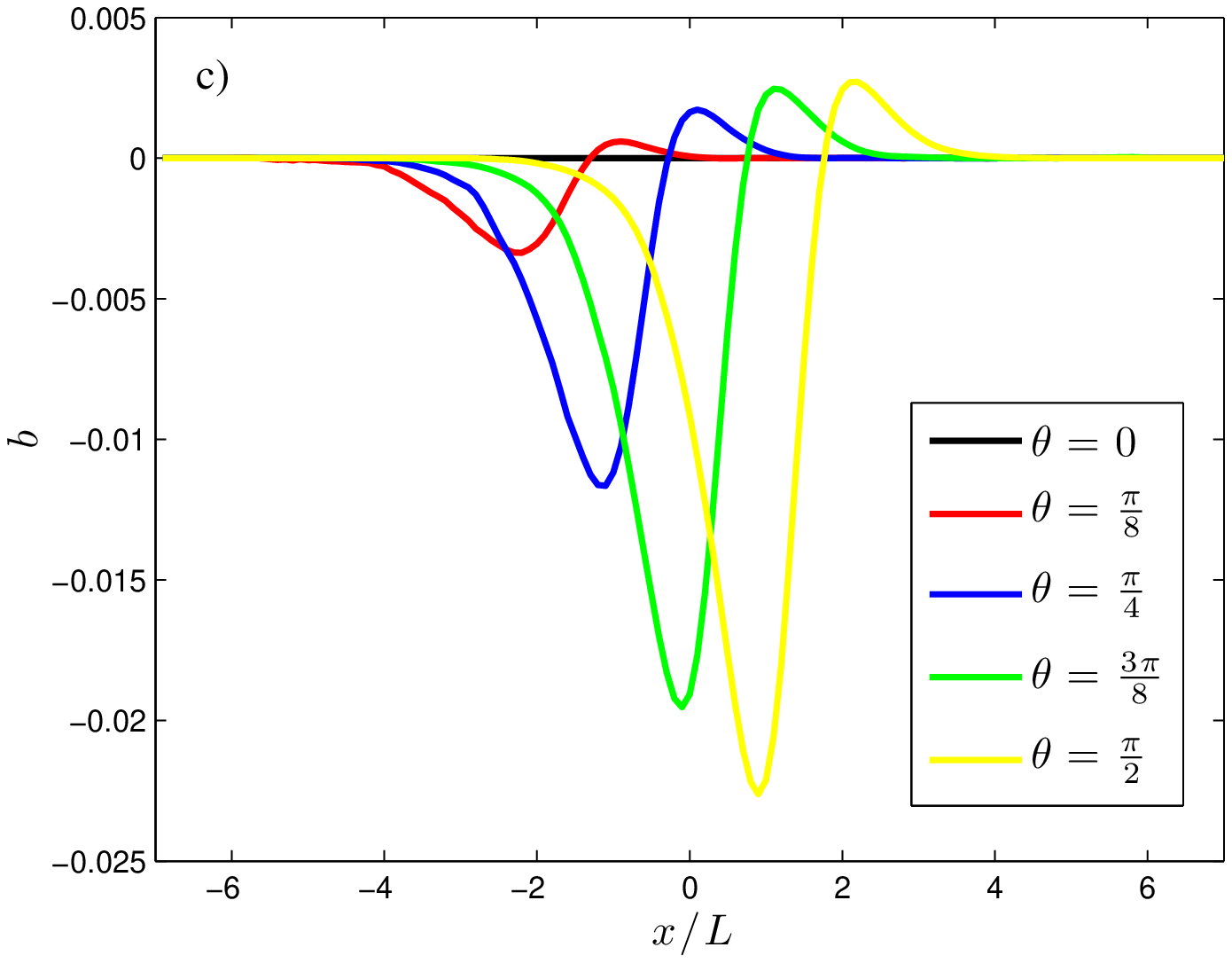}
\end{center}
\end{minipage}
\caption{The profiles of the order parameters: a) The density $c(x)$; b) The principal nematic order parameter $s(x)$; c) The biaxial effect $b(x)$. The definitions of $s(x)$ and $b(x)$ are given by denoting Q-tensor as $Q = s(\mathbf{nn} - \frac{1}{3} \mathbf{I})+b(\mathbf{n'n'} - \frac{1}{3}\mathbf{I})$, where $\mathbf{n'}$ is perpendicular to $\mathbf{n}$. The centers of these profiles are shifted by $0.5$ in order to clearly display them for different tilt angles. }\label{curve1}
\end{figure}

\reffig{curve1} (a) shows the profile of the density $c(x)$. The density profile for tilt angle $\theta$ greater than $\frac{\pi}{4}$ is a monotonically increasing function; nevertheless,  for the tilt angle smaller than $\frac{\pi}{4}$, it displays a shallow dip near the isotropic side of the interface, and a small oscillation near the nematic side of the interface. The shallow dip near the isotropic side was captured by Chen and Noolandi \cite{numerical molecular} in numerical simulation of Onsager's molecular theory, in which it was explained by the competition between the entropy and the excluded-volume interaction. The oscillating behavior was noticed by Allen \cite{interface mc} in Monte Carlo simulations; however, in Monte Carlo simulations, the oscillating behavior emerged for all cases of the tilt angle, and was explained as boundary effects. In the new Q-tensor model, the oscillating behavior can be explained by the negative coefficients of the derivative terms of density $c(x)$: with oscillation, the elastic energy $F_{elastic, 2}$ can be lower. On the contrary, the fourth order elastic energy $F_{elastic, 4}$ serves to stabilize the energy at the interface. The interfacial width of the density is roughly $3L$. It is at its narrowest at $\theta = \frac{\pi}{2}$. 

\reffig{curve1} (b) shows the principal nematic order parameter $s(x)$. Unlike the density profile, the profile of $s(x)$ is always a monotonic function. The centers of the profiles for $c(x)$ and $s(x)$ are different. There is a "phase shift" for $s(x)$ to the isotropic side with roughly $0.5L$ compared to the density profile. Unlike the density profile, the interfacial width of $s(x)$ remains identical at $3L$ as tilt angle $\theta$ increases.

The interfacial width is not sensitive to the coefficients $H_i$. No matter how we alter the coefficients $H_i$ with an order of magnitude, the interfacial width is always roughly $3L$; nevertheless, as $H_i$ increases, the oscillation near the nematic side is mitigated but still exists. 

\reffig{curve1} (c) shows the biaxiality parameter $b(x)$. The biaxial effect only appears significantly in the interfacial region. It is opposite between the isotropic side and the nematic side, and is stronger at the isotropic side. The biaxiality is the strongest for $\theta = \frac{\pi}{2}$. There is no biaxiality for the $\theta=0$ case because of the rotational symmetry. In general, the biaxial effect for $\pI$-$\pN$ interface is quite weak.

\begin{figure}
\begin{center}
 \includegraphics[width=0.55\linewidth]{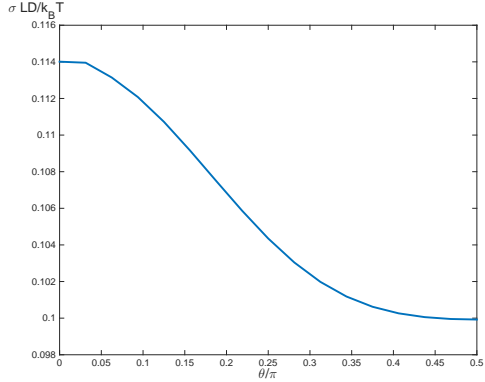}
\end{center}
\caption{The surface energy density versus the tilt angle $\theta$}\label{surface}
\end{figure}

\reffig{surface} shows the dependence of the isotropic-nematic surface energy on the angle $\theta$. The surface energy is defined as the difference between the grand potential of the interface and that of the isotropic phase. The surface energy is a monotonic function of the tilt angle and has a minimum at $\theta=\frac{\pi}{2}$ with value $\sigma = 0.100 k_BT/(LD)$. Therefore, the interface prefers to have a tilt angle of $\frac{\pi}{2}$, which is the configuration as \reffig{snapshots} (c). In another word, if we do not assume the tilt angle $\theta$ constant, but just anchor it at the boundary, it would tend to $\frac{\pi}{2}$ at the interface, which was the setup and situation in the study by Monte Carlo simulations \cite{interface mc}. 

There are several previous studies of the $\pI$-$\pN$ interface problem investigated using Onsager's molecular theory with volume exclusion potential for rod-like molecules \cite{numerical molecular} \cite{interfacial tension2} \cite{interface mc}. These simulations gave the interfacial width roughly at $2L$, which agree with the new Q-tensor model. Chen and Noolandi \cite{numerical molecular} predicted that the surface energy attains minimum at $\pi/2$ with value $\sigma= 0.183 k_BT/(LD)$. Koch and Harlen \cite{interfacial tension2} used a generalization of Onsager's approach and did a more comprehensive investigation. They predicted that the surface energy attains minimum at $\pi/2$ with value $\sigma= 0.316 k_BT/(LD)$. Our new Q-tensor model gives it a lower value $\sigma = 0.100 k_BT/(LD)$. Since the new Q-tensor model is a phenomenological model, and we alter the coefficients in our models, the value of the surface energy is only comparable in orders. To find an accurate value of the surface energy, it is better to use Onsager's molecular theory.

The experimental study of MBBA \cite{interfacetheta1} showed that the surface energy attains minimum at the tilt angle $\pi/2$, which is consistent with our calculations. However, the study of nCB \cite{interfacetheta2} showed that the minimum of surface energy is attained at a tilt angle close to  $\pi/3$. In the meanwhile, the experimental values of the interfacial width for the nCB are of the order of $400-700\mathring A$ (with molecule's length about $20-30\mathring A$), in which the ratio is much larger than the ratio in our work and numerical simulations with hard rod molecules \cite{numerical molecular} \cite{interfacial tension2} \cite{interface mc}. The numerical simulation of the $\pI$-$\pN$ interface problem in modeling specific materials is still under investigation.

\section{Comparison with Classical Models}\label{section: compare}

In the paper \cite{HLWZ}, the Oseen-Frank model and the Ericksen model were derived using the new Q-tensor model by assuming a constant density. In this section, we show the effectiveness of the new Q-tensor model by comparing it with other classical models in various scales aiming to model density variations. 

\subsection{Generalized Landau-de Gennes Models}

De Gennes pioneered the Q-tensor, which provides information on both the preferred molecular orientation and the degree of orientational order, to describe nematics phases \cite{model: de Gennes}. Later on, by exploiting an analogy between smectic liquid crystals and superconductors, de Gennes again proposed an independent, complex order parameter that allows for the description of nematic-smectic phase transitions \cite{model: de Gennes2}. Successive models generalized the Landau-de Gennes model and coupled the nematic order tensor with the complex smectic order parameter. In the work of Biscari \textit{et al.} \cite{model: B C T}, the free energy contained the polynomial terms and the first order derivative terms. In the work of Mukherjee \cite{model: Mukherjee}, the free energy contained the polynomial terms, the first order and the second order derivative terms of these order parameters. These models successfully modeled the $\pI$, $\pN$, $\pA$, and $\pC$ phases with computational efficiency. However, it is noteworthy that the complex order parameter is not suitable for numerical simulations of liquid crystals in confined geometries, such as disclinations and defects. The work of Pevnyi \textit{et. al.} \cite{PSS} addressed the above problem and replaced the complex order parameter with the density $c(\tx)$. The other order parameter $\ttn$ entered the free energy through the second order tensor $n_in_j$, so we regard this model as a tensor model. This model gave reasonable numerical results to model the $\pA$ phase in confined geometries with defects. The free energy contains the polynomial terms of the density, and the second order derivative terms of the order parameters, which can be viewed as a reduced case of the new Q tensor model. 

The new Q-tensor model are more generalized comparing to the above models. For the new Q-tensor model, the entropy term is in the thermotropic form defined by Ball and Majumdar \cite{paper: jball}, but no longer has the polynomial terms as in the Landau-de Gennes model. Such an entropy term ensures the boundedness of the eigenvalues of the Q-tensor, preserving its physical meaning. If we assume $c(\tx)$ is constant, $F_{bulk}+F_{elastic, 2}$ would degenerate to the energy functional of the Landau-de Gennes model, which is able to characterize the stable nematic phase \cite{paper: jball}.  

The introduction of $Q_4(\tx)$ is necessary, since otherwise the elastic constant $K_1$ would equal to $K_3$, which is not always true. Moreover, if we need to model density variation phenomena in which $c(\tx)$ is not constant, $F_{bulk}+F_{elastic, 2}$ would not be bounded from below, for the reason that the derivative terms of $c(\tx)$ in $F_{elastic, 2}$ are negative. It is necessary to consider at least the second order derivative terms $F_{elastic, 4}$ to stabilize the energy functional. 

\subsection{Generalized McMillan Models}\label{section: GMM}

Based on Onsager's molecular theory, assuming a two particle interaction potential and introducing an orientational order parameter, Maier and Saupe proposed a molecular field theory to characterize the $\pI$-$\pN$ transition \cite{model: Maier Saupe}. Adding a positional order parameter, McMillan extended the theory to include the $\pA$ phase \cite{model: McMillan}. Successive improvements were made to account for more complicated potentials, such as the work of Mederos \textit{et al.} \cite{model: molecular hard core} which incorporated the hard core interaction, and the work of Gorkunov \textit{et al.} \cite{model: molecular potential} which considered the fragments' attraction and repulsion interaction.  

Most recently, Pajak and Osipov \cite{model: P O} proposed a generalization of the McMillan model including $\mathbf{I}$-$\mathbf{N}$-$\mathbf{S_A}$-$\mathbf{S_C}$ phases. In the smectic phase, let $\mathbf{k}$ be the unit layer normal, $\mathbf{l}$ be the unit normal of the tilt plane, and $\mathbf{c}$ be their unit orthogonal complement. The layer thickness is presumed to be $2\pi/q$. The order parameters of the generalized McMillan model are $\tilde \psi$, $\tilde Q_{ij}$, and $\tilde \Sigma_{ij}$, defined in terms of the density function of molecules $f(\tx,\tm) = c(\tx)\rho (\tx,\tm)$ as
\begin{equation}
\begin{aligned}
\tilde \psi &= \frac{1}{\vert \Omega \vert}\int_\Omega  \int_{\mathbb{S}^2} f (\tx, \tm) \cos(q\mathbf{k}\cdot \tx)\ud \tm  \ud\tx, \\
\tilde Q_{ij} &= \frac{1}{\vert \Omega \vert}\int_\Omega \int_{\mathbb{S}^2} (\tm \tm - \frac{1}{3}\mathbf{I})f(\tx, \tm)\ud \tm \ud\tx, \\
\tilde \Sigma_{ij} & = \frac{1}{\vert \Omega \vert} \int_\Omega  \int_{\mathbb{S}^2} (\tm \tm - \frac{1}{3}\mathbf{I})f(\tx, \tm) \cos(q\mathbf{k}\cdot \tx) \ud \tm  \ud\tx.\\
\end{aligned}
\end{equation}

In the $(\mathbf{k}, \mathbf{c}, \mathbf{l})$ framework, these order parameters have components decomposition
\begin{equation}
\begin{aligned}
\tilde Q_{ij} &=  S(k_ik_j-\frac{\delta_{ij}}{3})+\frac{1}{2}P (c_i c_j -l_i l_j)+\frac{1}{2}V(k_i c_j +c_i k_j),\\
\tilde \Sigma_{ij} &=  \Gamma (k_ik_j-\frac{\delta_{ij}}{3})+\frac{1}{2}\Pi (c_i c_j -l_i l_j)+\frac{1}{2}\Lambda(k_i c_j +c_i k_j).\\
\end{aligned}
\end{equation}

The free energy of the generalized McMillan model can be expressed as
\begin{equation}
F=-\frac{1}{2}c_0 [u(S^2+\frac{3}{4}P^2 +\frac{3}{4}V^2) + \omega_0 \tilde \psi^2 + 2 \omega_1 \tilde \psi \Gamma + \omega_2 \Gamma^2 + \omega_3 \Pi^2 + \omega_4 \Lambda^2]-k_B T c_0 \ln Z,
\end{equation}
which represents the sum of the entropy and the quadratic terms of $\tilde \psi$, $S$, $P$, $V$, $\Gamma$, $\Pi$, and $\Lambda$. Here, $c_0$ is the average density, $u$ is the Maier-Saupe
constant which determines the $\pI$-$\pN$ transition temperature, $\omega_i$ are constants, and $Z$ is the partition function which can be represented by the order parameters.

The generalized McMillan model can be directly derived from the new Q-tensor model if we assume
\begin{equation}
\begin{aligned}
c(\tx)&=1+\tilde \psi \cos(q\mathbf{k}\cdot \tx),\\
c(\tx)Q_{ij}(\tx) &= \tilde Q_{ij} + \tilde \Sigma_{ij} \cos(q\mathbf{k}\cdot \tx).\\
\end{aligned}
\end{equation}
where $c(\tx)$ and $Q(\tx)$ are the order parameters of the new Q-tensor model, and $\tilde \psi$, $\tilde Q$, and $\tilde \Sigma$ are the order parameters of the generalized McMillan model. Intuitively, the order parameters of the generalized McMillan model are the Fourier coefficients of the order parameters of the new Q-tensor model, and the generalized McMillan model can be interpreted as the one mode approximation of the new Q-tensor model.

The generalized McMillan models are mechanistic models based on self-consistent field theory. The coefficients are linked with microscopic quantities, i.e. the parameters in interaction potentials. Another advantage of the generalized McMillan model is its elegance and simplicity, where the energy functional can be expressed by several scalar parameters. 

However, the order parameters of the generalized McMillan model are averaged over the entire space, failing to characterize some small scale phenomena where the domain has confined geometry and spatial variance, for example, phase transitions in confined geometries, wetting phenomena, surface-induced bulk alignment, defects and disclinations, and interfaces.

\subsection{Chen-Lubensky Model}\label{section: CL}

The Chen-Lubensky model \cite{model: chen} is a vector model based on the Landau-Ginzburg theory, characterizing the $\pN$-$\pA$-$\pC$ transitions. The order parameters are the density $c(\tx)$ and the director $\ttn(\tx)$. The corresponding free energy consists of two parts,
\begin{equation}
F_{CL} = F_c + F_{OF},
\end{equation}
where $F_c$ contains terms up to second order derivatives of the density $c(\tx)$,
\begin{equation}
\begin{aligned}
F_c=&\frac{1}{2\beta} \int_\Omega \Big(ac^2 + D_{1} \big[( \ttn \cdot \nabla)^2 c\big]^2
-C_{1}( \ttn \cdot \nabla c)^2+\frac{C_{1}^2}{4D_{1}}c^2 \\
&\qquad+C_{2}(\delta_{ij}-n_in_j) \nabla _i c \nabla_j c + D_2 [(\delta_{ij}-n_in_j)\nabla_i 
\nabla_j c]^2\Big) \ud\tx,
\end{aligned}
\end{equation}
and $F_{OF}$ is the Oseen-Frank energy for distortions in terms of the nematic director $\ttn(\tx)$,
\begin{equation}
\begin{aligned}
F_{OF}=\frac{1}{2}\int_\Omega  \Big(K_1 (\nabla \cdot  \ttn)^2 + K_2[ \ttn \cdot (\nabla \times  \ttn)]^2
 + K_3 \vert \ttn \times (\nabla \times  \ttn) \vert ^2\Big) \ud\tx.
\end{aligned}
\end{equation}
The coefficients $C_{1}$, $C_{2}$, $D_{1}$, $D_{2}$, $a$, and $K_i$ are determined through experiments. These coefficients are intuitive: $K_i$ are elastic constants; $C_{1}$ and $D_{1}$ determine the horizontal period; $C_{2}$ and $D_{2}$ determine the perpendicular period.

In the paper \cite{HLWZ} where the new Q-tensor model was first introduced, assuming the Q-tensor uniaxial and $c(\tx)$ and $S_2(\tx)$ constant, the Oseen-Frank energy $F_{OF}$ was deduced and the elastic constants $K_1$, $K_2$, $K_3$ were expressed analytically. In the following paragraphs, making alternative assumptions that Q-tensor is uniaxial and constant, we will deduce a vector model similar to the free energy $F_c$. 

Assume Q-tensor is uniaxial and spatially homogeneous,  
\begin{equation}\label{s2s4}
\begin{aligned}
Q_{ij} &= S_2 \left(n_in_j-\frac{1}{3} \delta_{ij}\right),\\
Q_{4ijkl} &= S_4 \left(n_in_jn_kn_l-\frac{1}{7}(n_in_j\delta_{kl})_{sym}+\frac{1}{35}(\delta_{ij} \delta_{kl})_{sym}\right),
\end{aligned}
\end{equation}
where $S_2$, $S_4$, and $\ttn$ are constant.  

Under this assumption, the energy functional can be formulated as
\begin{equation}\label{vector model}
\begin{aligned}
\beta F[c(\tx), S_2, \ttn] = & \int_\Omega c \ln (c) \ud \tx + \frac{1}{2}\int_\Omega \biggl(\hat A_1 c^2 -\hat G_1 \vert \nabla c \vert^2 - \hat G_2 (\ttn \cdot \nabla c)^2+ \hat H_1 (\nabla^2 c :\ttn\ttn)^2\\
& +\hat H_2 (\nabla^2 c:\ttn\ttn)\triangle c + \hat H_3 ( \triangle c )^2 +\hat H_4\vert \ttn \cdot \nabla^2 c \vert^2 + \hat H_5 \vert \nabla^2 c\vert^2 \biggl) \ud \tx,\\
\end{aligned}
\end{equation}
where $\hat A_i$, $\hat G_i$, and $\hat H_i$ are functions of $A_i$, $G_i$, $H_i$, $S_2$, and $S_4$, linear with respect to $A_i$, $G_i$, $H_i$, and quadratic with respect to $S_2$ and $S_4$. The expressions of $\hat A_i$, $\hat G_i$, $\hat H_i$ in terms of $A_i$, $G_i$, $H_i$, $S_2$, $S_4$ are listed in the appendix. 

The deduced vector model is able to characterize $\pI$-$\pN$-$\pA$-$\pC$ phase transitions. Let the average density be $c_0=\frac{1}{\vert\Omega\vert}\int_{\Omega} c(\tx) d\tx$, the entropy term is almost linear to $c_0$, and the interaction terms are quadratic to $c_0$. As $c_0$ is small, the entropy term dominates, and the system exhibits isotropic phases. As $c_0$ increases, the interaction terms dominate, and the system transits to nematic and smectic phases. 

Two quantities are important in smectic phases: layer thickness and tilt angle. To derive these quantities, assuming a given $S_2$, we plug a trial function $c( \tx) = c_0 + c_1 \cos( q \mathbf{k} \cdot  \tx)$ into the energy functional, where $\vert \mathbf{k} \vert =1$, $\mathbf{k} \cdot \ttn = \cos \theta$, and $\vert c_1\vert <1$. The free energy becomes
\begin{equation}
\begin{aligned}
\beta F(q^2, \cos^2 \theta, c_1)= &\int_{\Omega} c \ln (c) \ud \tx + \frac{1}{2}\int_{\Omega} \left\{ \hat A_1 c^2
 +  c_1^2 \cos^2(q\mathbf{k} \cdot \tx)\left[ - (\hat G_1  + \hat G_2 \cos^2 \theta)  q^2\right. \right.\\
&  \left.\left.+(\hat H_1 \cos^4 \theta + \hat H_2 \cos^2 \theta+\hat H_3 +\hat H_4 \cos^2 \theta +\hat H_5)  q^4 \right]\right\}\ud\tx,
\end{aligned}
\end{equation}
where $F$ is a quadratic function of $q^2$ and $\cos^2 \theta$. Minimizing the free energy over $q$, $\cos \theta$, and $c_1$, we obtain the minimizer $\hat q$, $\hat \theta$, and $\hat c_1$, with relations
\begin{equation}
\begin{aligned}
\hat q^2 &= \frac{\hat G_1 +\hat G_2 \cos^2 \hat \theta}{2(\hat H_1 \cos^4 \hat \theta
+ \hat H_2\cos^2 \hat \theta + \hat H_3 + \hat H_4 \cos^2 \hat \theta + \hat H_5)},\\
\cos^2 \hat \theta &= 1-\frac{ \hat q^2(2 \hat H_1+ \hat H_2+\hat H_4)-\hat G_2}{2  \hat q^2 \hat H_1}.
\end{aligned}
\end{equation}
These two equations can be solved iteratively, and $\hat c_1$ is easy to compute but cannot be expressed explicitly. The system will exhibit
\begin{itemize}
\item $\pA$ phase, if $\hat q^2 > 0$, $\cos^2 \hat \theta \geq 1$, and $\hat c_1 \neq 0$.  Let $\hat q_1 = (\hat G_1 +\hat G_2)/[2(\hat H_1 + \hat H_2 + \hat H_3 + \hat H_4 + \hat H_5)]$. The layer thickness is $\hat h=2\pi L/\hat q_1 $.
\item $\pC$ phase, if $\hat q^2 >0$, $0 < \cos^2 \hat \theta <1$ , and $\hat c_1 \neq 0$. The layer thickness is $\hat h = 2\pi L/\hat q$, and the tilt angle is $\hat \theta$.
\item no smectic phase, otherwise.
\end{itemize}

The analytical derivation above can explain the relations between the coefficients $A_i$, $G_i$, $H_i$, and the tilt angle and layer thickness. It can also help to set the coefficients in the new Q-tensor model.

In comparison with the deduced vector model, the Chen-Lubensky model lacks entropy terms which leads to its failure in characterizing the isotropic phase. In the Chen-Lubensky model, the Oseen-Frank energy and the derivative terms of $c(\tx)$ are added up directly. On the contrary, in the new Q-tensor model, the Oseen-Frank energy and the derivative terms of $c(\tx)$ are deduced based on different assumptions as discussed previously. If no assumptions are made, there will be cross terms incorporating the gradient of $\ttn$ and the gradient of $c(\tx)$. Furthermore,  there are two extra terms in the deduced vector model, $\hat H_4 \vert n \cdot \nabla^2 c\vert^2$ and $\hat H_5 \vert \nabla^2 c \vert^2$. These two terms may lead to different phenomena in the subspace perpendicular to the director. Except for these differences, the deduced vector model agrees with the Chen-Lubensky model.

\section{Summary}\label{section: sum}

In this work, we have investigated the new Q-tensor theory for liquid crystals focusing on density variations. Phenomenologically, the $\pI$-$\pN$-$\pA$-$\pC$ phase transitions are predicted, and the $\pI$-$\pN$ interface is investigated. We have also drawn comparisons of the proposed model with the classical models, and strong connections are found. We have shown that all these classical models can be derived from the new Q-tensor model.

In comparison, the high dimension of the order parameter sets considerable obstacles in computations for the Onsager's molecular model. The order parameter $\ttn(\tx)$ adopted in Chen-Lubensky model fails to explain the degree of orientational order. Landau-de Gennes model is famous for its tractability in mathematical analysis and computations. However, generalizations of Q-tensor model are needed to include smectic phase. The generalized McMillan model is a good approximation of the Onsager's theory, and efficient in computations. The weakness is in the ultimate energy functional which fails to reflect detailed physical understanding. In addition, the generalized Landau-de Gennes models did not explain how their energy functionals were derived in details.

On the other hand, the new Q-tensor model can be derived from Onsager's molecular theory. The introduction of density enables us to characterize smectic phases and the density variations in physical phenomena. The new Q-tensor model captures much of the essential physics while remaining mathematically tractable and efficient computationally. Compared to all classical models, the new Q-tensor model is a good trade-off.

One notes that the coefficients $A_i$, $G_i$, $H_i$ in the new Q-tensor model cannot be related with any specific liquid crystal materials so far. The physical relevance of the coefficients is still under investigation. The new Q-tensor model studies the phase transitions as a function of the average density. It is difficult to consider temperature-driven phase transitions, for which one needs to include the attraction interaction. This will be studied in future work. Defects are essential physical phenomena for liquid crystals. The effects of density variations in defects characterized by the new Q-tensor model will also be studied in the future.

\section*{Acknowledgements}

{\small
The authors wish to thank Dr. Jiequn Han and Dr. Wei Wang for their help in discussions and advices for writings. Pingwen Zhang is partly supported by NSF of China under Grant 11421101 and 11421110001.
}

\Appendix
\section*{}

The relations between $\hat A_i$, $\hat G_i$, $\hat H_i$ (the coefficients in vector model with energy functional \refequ{vector model}) and $A_i$, $G_i$, $H_i$ (the coefficients in the new Q-tensor model with energy functional \refequ{energy0}) are as follows. $S_2$ and $S_4$ are the principal nematic order parameters defined as \refequ{s2s4}.

\begin{equation}
\begin{aligned}
\hat A_1 &= A_1-\frac{2}{3}A_2 S_2^2 -\frac{8}{35}A_3 S_4^2\\
\hat G_1 &= G_1 -\frac{2}{3}S_2^2G_2 -\frac{2}{9}S_2^2 G_3 - \frac{1}{3}S_2 G_4-\frac{8}{35}S_4^2 G_5-\frac{12}{245}S_4^2 G_6 + \frac{4}{35} S_2S_4G_7\\
\hat G_2 &= -\frac{1}{3}S_2^2 G_3 + S_2G_4-\frac{4}{49}S_4^2 G_6-\frac{12}{35}S_2 S_4 G_7\\
\hat H_1 &= S_2^2 H_3\\
\hat H_2 &=  -\frac{2}{3} S_2^2 H_3\\
\hat H_3 &=  \frac{1}{9}S_2^2 H_3\\
\hat H_4 &= \frac{1}{3} S_2^2 H_4\\
\hat H_5 &= H_1+\frac{2}{3}S_2^2 H_2+\frac{1}{9}S_2^2 H_4\\
\end{aligned}
\end{equation}

\end{document}